\documentclass[pdftex,twocolumn,aps,prd,superscriptaddress,groupedaddress]{revtex4}          

\RequirePackage[T1]{fontenc}

\RequirePackage{amsmath}
\RequirePackage{graphicx}
\RequirePackage{mathptmx}      
\RequirePackage{flushend}
\RequirePackage[colorlinks,citecolor=blue,urlcolor=blue,linkcolor=blue]{hyperref}

\begin{document}
\newcommand{\EEG}{\rm e^+ e^-\rightarrow \gamma\gamma}
\newcommand{\EEGG}{\rm e^+ e^-\rightarrow \gamma\gamma(\gamma)}
\newcommand{\EEGGG}{\rm e^+ e^-\rightarrow \gamma\gamma\gamma}
\newcommand{\EEGGGG}{\rm e^+ e^-\rightarrow \gamma\gamma\gamma\gamma}
\newcommand{\EEEEG}{\rm e^+ e^-\rightarrow e^+e^-(\gamma)}
\newcommand{\EEEE}{\rm e^+ e^-\rightarrow e^+e^-}
\newcommand{\EEmmtt}{\rm e^+ e^-\rightarrow \mu ^{+}\mu ^{-},\rm e^+ e^-\rightarrow \tau ^{+}\tau ^{-}}
\newcommand{\EEllgg}{\rm e^+ e^-\rightarrow \ell^{+}\ell^{-}\gamma\gamma}
\newcommand{\EEg}{\rm e^+ e^-\rightarrow 3\gamma}
\newcommand{\LAMP}{ \Lambda_{+}}
\newcommand{\LAMM}{ \Lambda_{-}}
\newcommand{\LAMS}{ \Lambda_{6}}
\newcommand{\LAMPP}{ \Lambda_{++}}
\newcommand{\LAMMM}{ \Lambda_{--}}
\newcommand{\DSDW}{ \sigma(\theta)}
\newcommand{\MESTAR}{ m_{{\rm e^{\ast}}}}
\newcommand{\EELL}{\rm e^+ e^-\rightarrow l^{+}l^{-}}
\newcommand{\Lagr}{\mathcal{L}}

\title{Hint for a minimal interaction length in $ \EEG $ annihilation in total cross section of centre-of-mass energies 55 - 207 GeV }

\author{Yutao Chen}\affiliation{Department of Modern Physics, University of Science and Technology of China, Jinzhai Road 96, Hefei, Anhui, 230026, China}
\author{Minghui Liu}\affiliation{Department of Modern Physics, University of Science and Technology of China, Jinzhai Road 96, Hefei, Anhui, 230026, China}
\author{J\"{u}rgen Ulbricht}\affiliation{Swiss Institute of Technology ETH Zurich, CH-8093 Zurich, Switzerland}



\begin{abstract}
The measurements of the total cross section of the $ \EEGG $ reaction from the VENUS, TOPAS, OPAL, DELPHI, ALEPH and L3 collaborations,
collected between 1989 to 2003, are used to perform a $ \chi^{2} $ test to search for a finite interaction length in direct contact term.
The experimental data of the total cross section compared to the QED cross section of a $ \chi^{2} $ test allows, to set a limit on a finite interaction length $ r_{e}=(1.25\pm 0.16) \times 10 ^{-17} [cm] $. In the direct contact term 
annihilation is this interaction lengths a measure for the size of the electron.
\end{abstract}

\maketitle

\section{Introduction}

The Standard Model of particle physics ( SM ) is the theory describing, three of the four known fundamental forces 
(the electromagnetic ( QED ), weak, and strong interactions ). After the discovery of the three families of fundamental particles, 
the gluons , photon, Z - boson and W - boson, the discovery of the Higgs in 2012 established the last cornerstone of the ( SM ).
Missing so far is a unification between SM and the fours interaction the gravitation. It is for this reason 
essential  to investigate  deviations of the SM. All the three fundamental forces have a definite interaction length.
In particular the QED has an infinite interaction length. In this paper we search for minimal interaction length. A minimal
interaction length would be an indication of a deviation from the QED. The concept of a minimal 
interaction length suggested by path - integral quantisation \cite{Planck},   
string theory \cite{String1,String2}, black hole physics \cite{Black1}, and quantum gravity \cite{QG:Qgrav1, QG:Qgrav2, QG:Qgrav3} has been introduced
into quantum mechanics and quantum field theory through generalised uncertainty principle which restricts an
accuracy of $ \Delta l $ in measuring a particle position by a certain finite
minimal length scale $ l_{m} $ related to maximum resolution \cite{RES:RES1, RES:RES2, RES:RES3} ( for a review \cite{MLI} ) .
In gravity, the limiting quantum length is the Planck length 
$ l_{P}=\sqrt{\hbar G/c^{^{3}}} =1.6\times 10^{^{-33}} $ cm, the related energy scale $ M_{P}\simeq 10^{16} $ TeV.
However gravitational effects have only been tested up to 1 TeV scale \cite{PG11} which corresponds to 
$ l_{m}\simeq 10^{-17} $ cm \cite{SMS}; therefore, minimal length could be in principle found within the range 
$ l_{P} $ and  $ l_{m} $ \cite{SMS}. In models with extra dimensions, the Planck length can be reduced to $ 1/M_{f} $ 
with $ M_{f}\simeq 1 $ TeV, which results in modification of cross-sections of basic scattering processes 
\newline
$ {\EEmmtt} $ (\cite{PLRR} and references therein).

 In this paper, we summarise the results on investigating experimental data on the annihilation reaction 
 $ {\EEGG} $ motivated by search for manifestation of the non-point-like behaviour of fundamental particles.
 
 In a first approach, the question of intrinsic structure of a charged spinning particle like an electron has been discussed
 in the literature since its discovery by Thomson in 1897. In quantum field theory, a particle is assumed to be point-like,
 and classical models of point-like spinning particles describe them by various generalisations of the classical
 Lagrangian $ (-mc\sqrt{\dot{x}\dot{x}}) $ 
 \cite{CL:1,CL:2,CL:3,CL:4,CL:5,CL:6,CL:7,CL:8,CL:9,CL:10,CL:11,CL:12,CL:13,CL:14,CL:15,CL:16}. 
 Another type of point-like models 
 \cite{PL:1,PL:2,PL:3,PL:4,PL:5,PL:6,PL:7,PL:8,PL:9,PL:10,PL:11}
 goes back to the Schr\"{o}dinger suggestion that 
 the electron spin can be related to its Zitterbewegung motion \cite{Schrodinger}.
 
A second approach, works with extended particle models. Early electron models based on the concept 
of an extended electron, proposed by Abraham more than a hundred years ago 
\cite{Abraham,Lorentz}, encountered the
problem of preventing an electron from flying apart under the Coulomb repulsion. Theories based on 
geometrical assumption about the " shape" or distribution of a charge density were compelling to
introduce cohesive forces of non electromagnetic origin ( the Poincar\'{e}stress ) 
\cite{Dirac}. A new review of models is discussed in \cite{Burinskii}. In this paper we apply to this model
\cite{Burinskii}. To find evidence for an extended particle picture, we used 
available data of experiments performed to search for a non-point-like behaviour, which focus on
characteristic energy scales related to characteristic length of interaction region in reference 
\cite{INTA:1,INTA:2,INTA:3,INTA:4,INTA:5,INTA:6} . Experimental 
limits on size of a lepton in reference 
\cite{INTA:1,INTA:2,INTA:3,INTA:4,INTA:5} ,
appear to be much less than its classical radius which suggests the
existence of a relatively small characteristic length scale related to gravity in reference
\cite{INTA:1,INTA:2,INTA:3,INTA:4,INTA:5}. 

To investigate the pure electromagnetic interaction 
\newline
$\EEGG$ using differential cross data from VENUS, TOPAS, ALEPH, DELPHI, L3 AND OPAL sets the limit on 
maximal resolution at scale E = 1.253 TeV by the character length $ l_{e}\simeq 1.57\times 10^{-17} $ cm
with a 5$\sigma $ significance \cite{INTA:6}. An earlier report set a 2.6$\sigma $ 
on an axial-vector contact interaction in the data on $ \EEEEG $ at centre - of - mass - energies 192 - 208 GeV \cite{Bourilkov}. 

\begin{figure}[htbp]
\vspace{0.0mm}
\begin{center}
 \includegraphics[width=7.0cm,height=4.0cm]{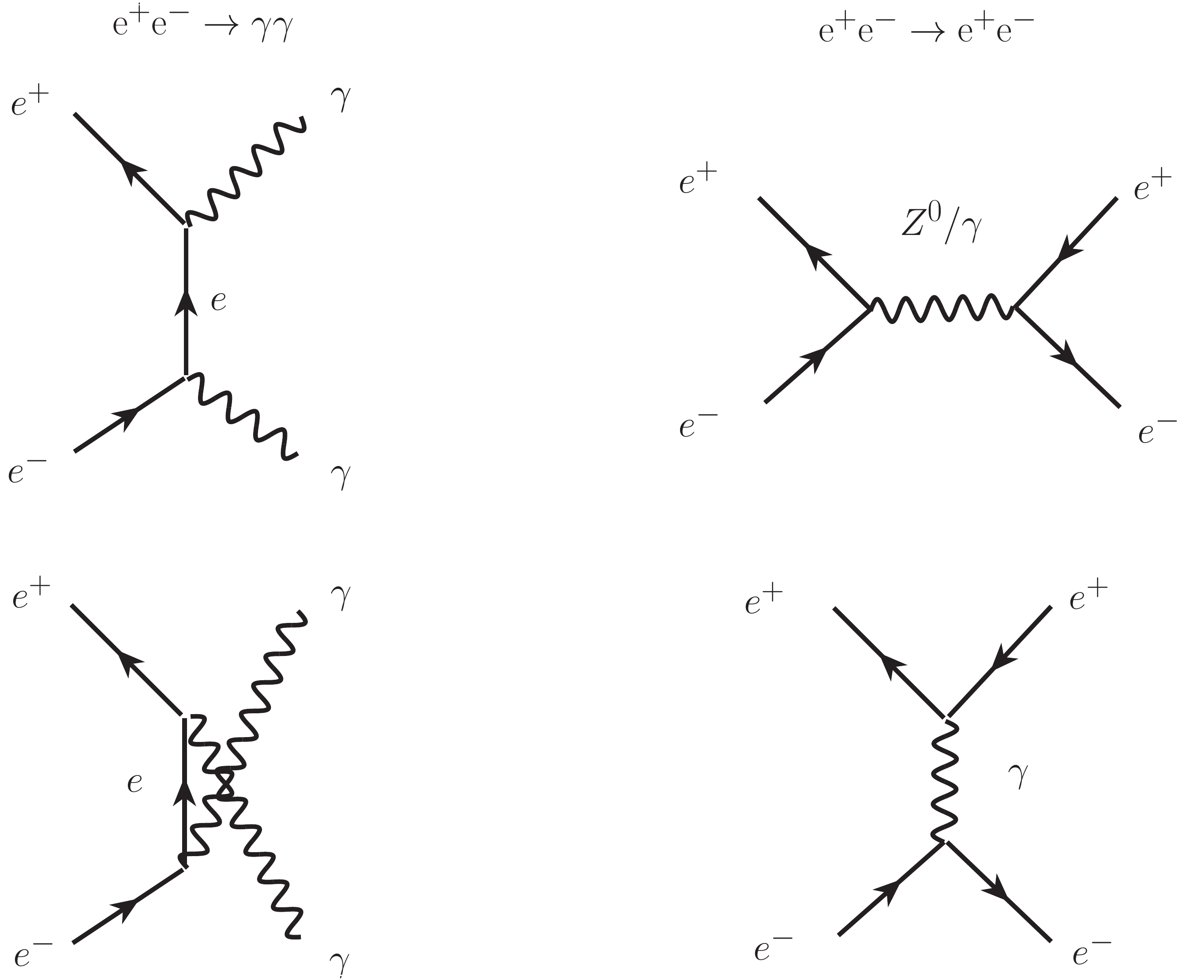}
\end{center}
\caption{ Lowest order $ \EEG $ and $ \EEEE $ reactions. The reaction $ \EEG $ proceeds via the t - and u - channel. 
The Bhabha $ \EEEE $ channel proceeds via the s - and t - channel. }
\label{Feynman1}
\end{figure}

The available data from the $ e^+ e^- $ accelerators favour two experiments  to test the finite interaction length 
the $ \EEG $ and $ \EEEE $ reactions. Both reactions are shown in Figure~\ref{Feynman1}.

The QED reaction $ \EEG $ is testing the behaviour of the electrodynamic long range force of the $ e^+ $  $ e^- $ reaction. 
The two $ \gamma $'s in the final state of the reaction $ \EEG $ are undistinguishable. The reaction performs for this reason via
the t - and u - channel. The s - channel is forbidden, by the law of angular momentum conservation. The two $ \gamma $'s in the final state are
left-handed and right-handed polarised. They couple to total spin zero. Under these circumstances the s - channel with 
spin one for $ \gamma $ and $ Z^0 $ is highly suppressed. 

The Bhabha reaction $ {\EEEE} $ is not only sensitive to the long range force of the electromagnetic $ e^+ $  $ e^- $ reaction, in addition 
via the $ Z^0 $ also to the short range force of the electroweak interaction. The reaction proceeds,  via scattering in the s - channel and
t - channel. The $ e^+ $ and $ e^- $ in the initial state and final state are identical. The gammas in the final state of the $ \EEG $ reaction 
disappear. The high charge sensitivity of the involved detectors allows to suppress the background $ {\EEEE} $ reaction, even under the
circumstances that at the  $ Z^0 $ pole the total cross section of the $ {\EEEE} $ reaction a factor two bigger as the 
total cross section of the $ \EEG $ reaction.

After the commissioning of the high energy $ e^+ e^- $ accelerators 1986 TRISTAN at KEK, the VENUS collaboration initiated 1989 the first experiments at 
$ \sqrt{s} $ = 55 GeV to investigate the total and differential cross section of the $ \EEG $ reaction. The experiments continued until LEP 
was closed 2000 at $ \sqrt{s} $ =  207 GeV.

In detail, the reaction was investigated by the VENUS \cite{VENUS} collaboration 1989 from energies $ \sqrt{s} $ = 55 GeV - 57 GeV, OPAL
\cite{OPAL1} collaboration 1991 at  the $ Z^0 $ pole at $ \sqrt{s} $ = 91 GeV, TOPAS \cite{TOPAS} collaboration 1992 at $ \sqrt{s} $ = 57.6 GeV, 
ALEPH \cite{ALEPH} collaboration 1992 at the $ Z^0 $ pole $ \sqrt{s} $ = 91.0 GeV, DELPHI \cite{DELPHI} collaboration 
from 1994 to 2000 at energies $ \sqrt{s} $ = 91.25 GeV to 202 GeV, L3 \cite{L3A} collaboration from 1991 to 1993 at the 
$ Z^0 $ pole range from $ \sqrt{s} $ = 88.5 GeV - 93.7 GeV, L3 \cite{L3B} collaboration 2002 from $ \sqrt{s} $ = 183 GeV - 207 GeV
 and OPAL \cite{OPAL2} collaboration 2003 from $ \sqrt{s} $ = 181 GeV - 209 GeV.

\begin{figure}[htbp]
\vspace{0.0mm}
\begin{center}
 \includegraphics[width=8.0cm,height=4.0cm]{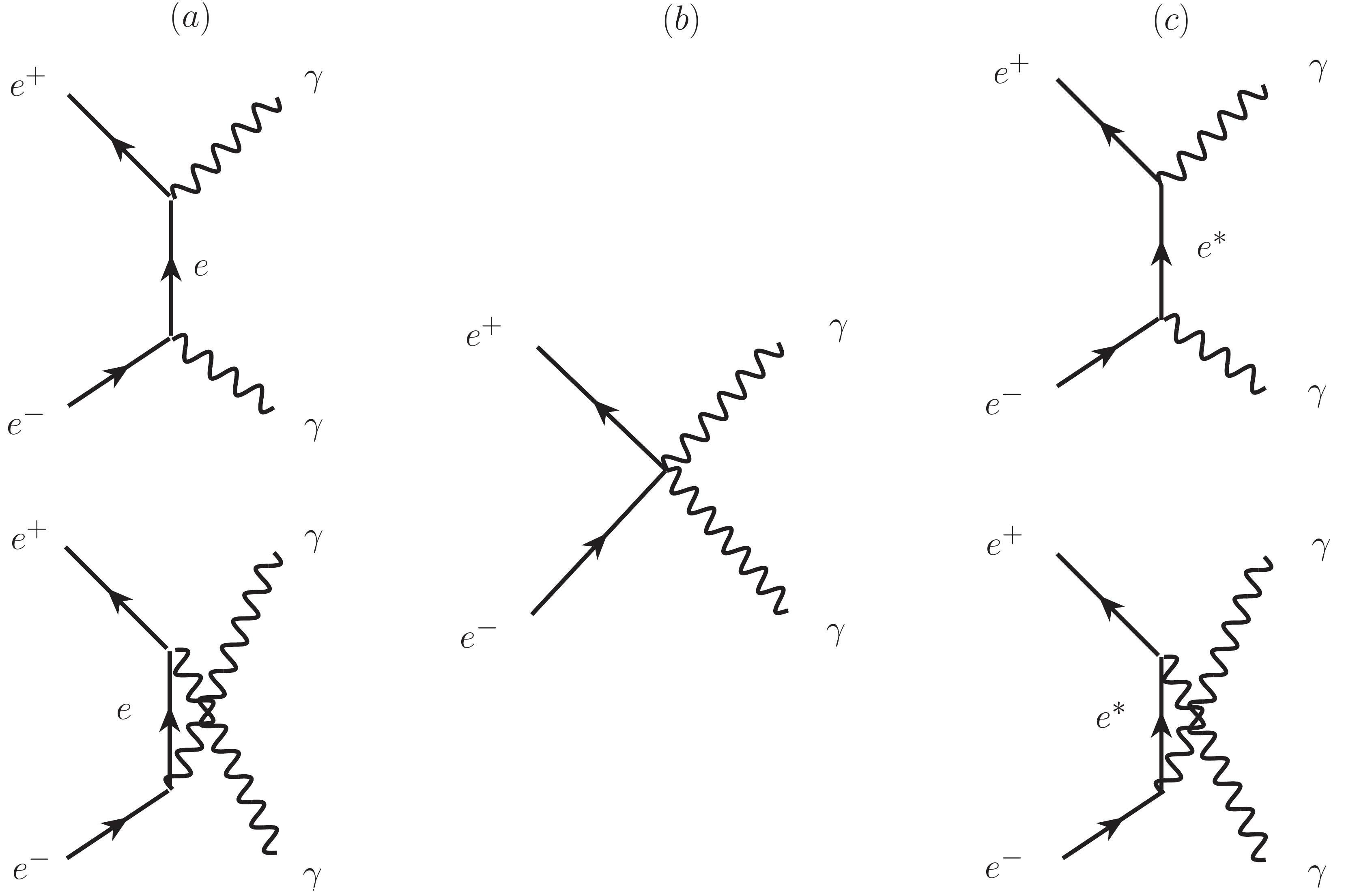}
\end{center}
\caption{ Lowest-order Feynman diagrams of $ \EEG $ reaction.  Figure ( a ) represents QED, 
                ( b) contact interaction and ( c ) excited electron exchange. }
\label{Feynman2}
\end{figure}

The experimental data of the differential cross section of these six collaborations from $ \sqrt{s} $ = 55 GeV to 207 GeV are
compared to the theoretical predicted QED differential cross section. Possible deviations from QED  were studied in terms 
of contact interaction and excited electron exchange shown in Figure~\ref{Feynman2}.

The deviation to the QED reaction is visible in the angular distribution of the gammas
from the experiment to the QED theory and in the experimental total cross section data to the QED. 
It is necessary to perform for example a $ \chi ^2 $ test or similar statistical tests, to search in the test for a minimum or limit for 
a scale parameter $ \Lambda $ [ GeV ]. This parameter finally allows to define a finite radius of the electron or the mass of an excited heavy
electron. It is important to notice, that a signal of deviation from QED is not visible in new final state particles like for the HIGGS search, it is 
hidden only in the angular distribution of the differential cross section and in the energy dependence of the total cross section to 
the QED values.

All the collaborations search for bounds on effective interactions from the reaction $ \EEG $  for example \cite{Models2}. Cut off
parameter $ \Lambda $ are used to set mass scales of different dimensionally interactions. Data of differential cross 
section are used to set limits on compositeness scales $ \Lambda_{+} $ and $ \Lambda_{-} $  in the direct 
contact interaction of the diagram shown in Fig. 2b and search for excited electrons $ m_{e^{*}} $ in the t, u-channel of the 
diagram shown in Fig. 2c. For example the L3 collaboration published four papers 1992, 1996, 1997 and 2000 and set
 ( 1992 ) \cite{L33}  limits on $ \Lambda_{+} $ > 139 GeV , $ \Lambda_{-} $ > 108 GeV and a limit on the mass of
 an excited electron to $ m_{e^{*}} $ >  127 GeV. 

\subsection{\it Outline of this paper}

The VENUS, TOPAS, OPAL, DELPHI, L3 and ALEPH  collaboration used the differential cross section of the
$ \EEG $ reaction to search for a deviation from QED. This was performed for certain energy ranges and luminosities. 
We perform a global $ \chi ^2 $ FIT, using the data from these six research projects to investigate $ \Lambda_{+} $, $ \Lambda_{-} $ 
and $ m_{e^{*}} $ for energies from $ \sqrt{s} $ = 55 GeV to 207 GeV including the associated luminosities \cite{QED-1}.
This analysis allowed to set an approximately 5$\times \sigma$ limit on the finite sizeof the electron $ r=(1.57\pm 0.28) \times 10 ^{-17} [cm] $
and  on the mass of an exited electron of $ m_{e^{*}}  = 308 \pm 56 [ GeV ] $. The deviation in the differential experimental
cross section from the QED values of approximately 4 $ \% $ was only visible in the fit results but not direct in the comparison
of the experimental and theoretical QED cross section.

The aim of this instigation is to prove in a $ \chi ^2 $ FIT, that the use of only the total cross section of all these data implies 
a similar result. First it is necessary to discuss the theoretical framework of the calculation of the total QED cross section,
in particular the fact that an analytic precise QED cross section must be calculated via a Monte Carlo program. Second, 
as no total cross section for the QED $ \EEG $ reaction from $ \sqrt{s} $ = 55 GeV to 207 GeV exists, we introduced a 
model for a total cross section using the data from the 6 collaborations. Including this information, it is possible to 
perform a total $ \chi ^2 $ FIT of all data. Finally it is possible to discuss the results.

\section{Theoretical framework}

To test the point structure of the electron, requests a high precision of the theoretical QED calculation of the differential 
and total cross section of the $ \EEG $ reaction.

The interactions of particles are dictated by symmetry principles of local gauge invariance and conserved physical quantities. 
Mathematical the Lagrangian formalism is used to connect symmetries and conservation laws. 

The Euler Lagrange equation is used to describe a free particle with spin 1/2 . In the minimum of the action path integral
is the Lagrangian density of the Dirac equation (\ref{DIRAC}).

\begin{equation}
\label{DIRAC}
   {\Lagr}_{Dirac}=\bar{\Psi }(i\gamma^{\mu}\partial _{\mu }-m)\Psi 
\end{equation}

With $ \Psi $ the fermion field, $ \bar{\Psi }\equiv \Psi ^{+}\gamma ^{0} $ its adjoint spinor, $ \gamma ^{\mu} $
the gamma matrices, $ \partial ^{\mu }=\partial /\partial x_{\mu } $ the covariant derivative and $ m $ the mass
of the particle.

A particle with interaction, is described by the local gauge invariance QED Lagrangian function in (\ref{QED1}).

\begin{equation}
\label{QED1}
   {\Lagr}_{QED}=\bar{\Psi }(i\gamma^\mu \partial _\mu -m)\Psi +e\bar{\Psi }\gamma ^\mu A_\mu \Psi -\frac{1}{4}F_{\mu \nu }F^{\mu \nu }
\end{equation}

With $ A_{\mu} $ the gauge field, the mass $ m_{A} = m_{ \gamma } = 0 $ , $ e $ charge of the electron,
 $ e\bar{\Psi }\gamma ^\mu A_\mu \Psi $ the interaction term and 
$ F_{\mu\nu}=\partial _{\mu}A_{\nu}-\partial _{\nu}A_{\mu} $. 
 
\subsection{\it The lowest order cross section of  $ {\EEG} $}

The interaction term of the QED Lagrangian allows to calculate the lower case Born cross section of the $ {\EEG} $ reaction.
The mathematical formalism uses the $ \it {M} $-matrix of (\ref{QED2})

\begin{equation}
\label{QED2}
M_{fi}=-e\int \bar{\Psi }\gamma ^{\mu }\Psi A_{\mu }d^{4}x
\end{equation}

Where, ( $ f $ ) stands for final and ( $ i $ ) for initial state. 
The derivation of the differential cross section uses the square of the $ \it {M} $-matrix, including the t - and u - channels of the 
Feynman graphs of Figure 2a and neglecting the electron mass $ m_{e} \approx 0 $ for high energies (\ref{BORN}).

\begin{equation}
\label{BORN}
\frac{d\sigma _0}{d\Omega }=\frac{S}{64 \pi ^2 s}\frac{p_f}{p_i}\left | \it {M}\right |^2=\frac{\alpha ^2}{s}\frac{1+cos^2(\theta  )}{k - cos^2(\theta )} 
\end{equation}

$ S = 1/2 $ is a statistical factor, $ \sqrt{s} $ the centre-of-mass energy, the momentum $ p_{f} = p_{i} $ , $ k=E_{e^+}/\left | \vec{p}_{e^+}\right |\simeq 1 $ 
for high energies $ E_{e^+} $ and $ \alpha = e^{2} / 4 \pi $. The angle $ \theta $ is the photon scattering angle with respect to the $ e^+ e^- $ - beam axis.

The total Born cross section is the integral over the angle $ \theta $ and the azimuth angle $ \phi $ ( \ref{BORN1} )

\begin{align}
\label{BORN1}
\sigma ^0&=\frac{1}{2!}\frac{\alpha ^2}{s}\int_{0}^{2\pi }d\phi \int_{-1}^{+1}\frac{1+cos^2\theta }{k-cos^2\theta }d(cos\theta ) & \vspace{0.5cm} \\ 
 &=\frac{2\pi \alpha ^2}{s}(ln(\frac{s}{m^2_{e}})-1) \nonumber 
 \end{align}

The precision of the BORN cross sections is absolute not sufficient to search for non-point-like behaviour of the electron in a $ \chi ^2 $ fit of 
the $ \EEGG $ reaction.

\subsection{\it Radiative corrections of the QED $ \EEGG $ cross section.}

All 6 collaborations used radiative corrections of the QED $ \EEGG $ cross section for virtual, soft and 
hard photons \cite{MC-tot}. In total 22 corrections are implemented \cite{Berends}.

\begin{figure}[htbp]
\vspace{0.0mm}
\begin{center}
 \includegraphics[width=7.0cm,height=6.0cm]{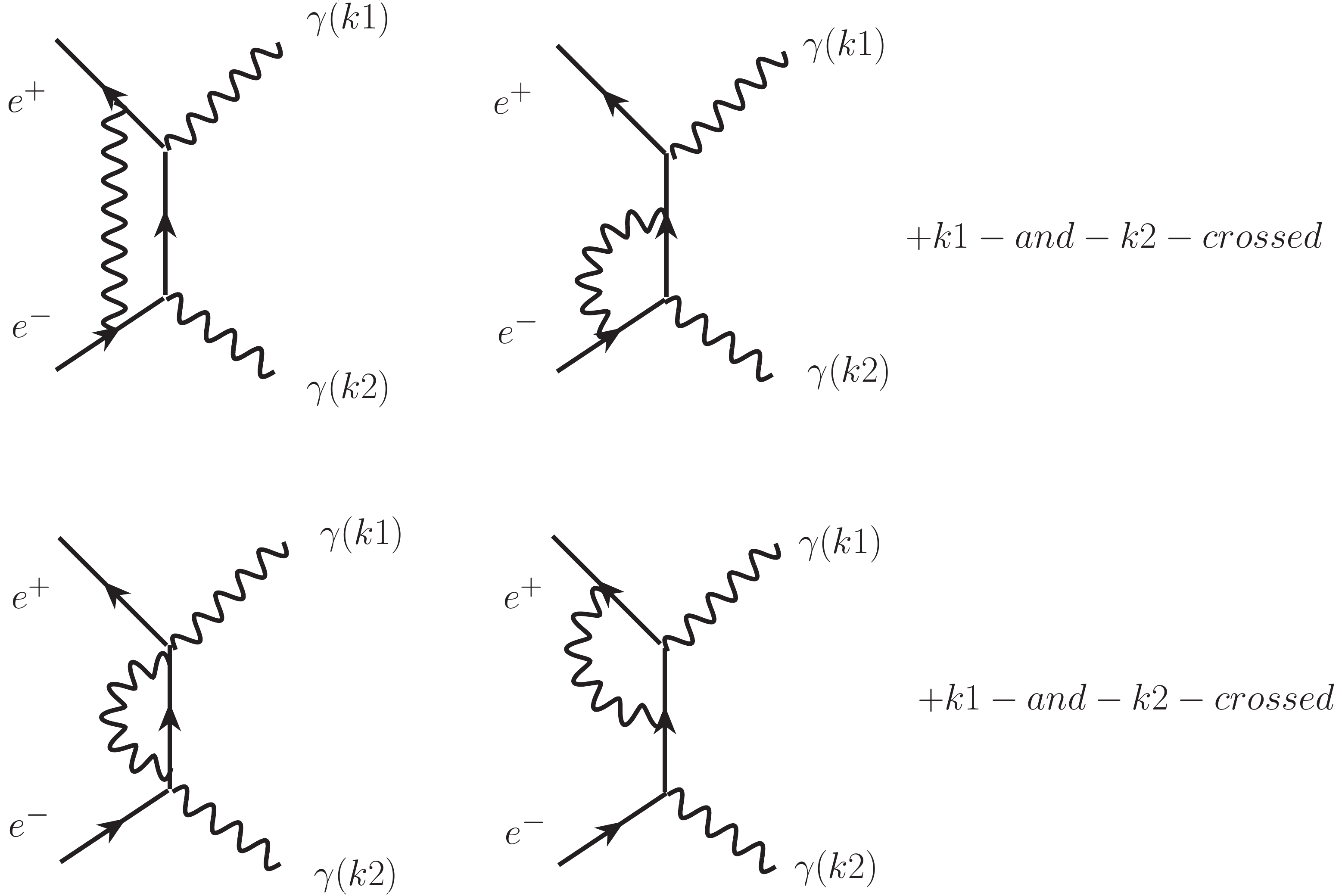}
\end{center}
\caption{ Third order of eight virtual photon corrections Feynman graphs of the $ \EEGG $  reaction. }
\label{CORR1}
\end{figure}

The first set of eight virtual photon corrections is shown in the Feynman graphs of Figure~\ref{CORR1} \cite{Fin-graph}.

\begin{figure}[htbp]
\vspace{0.0mm}
\begin{center}
 \includegraphics[width=7.0cm,height=6.0cm]{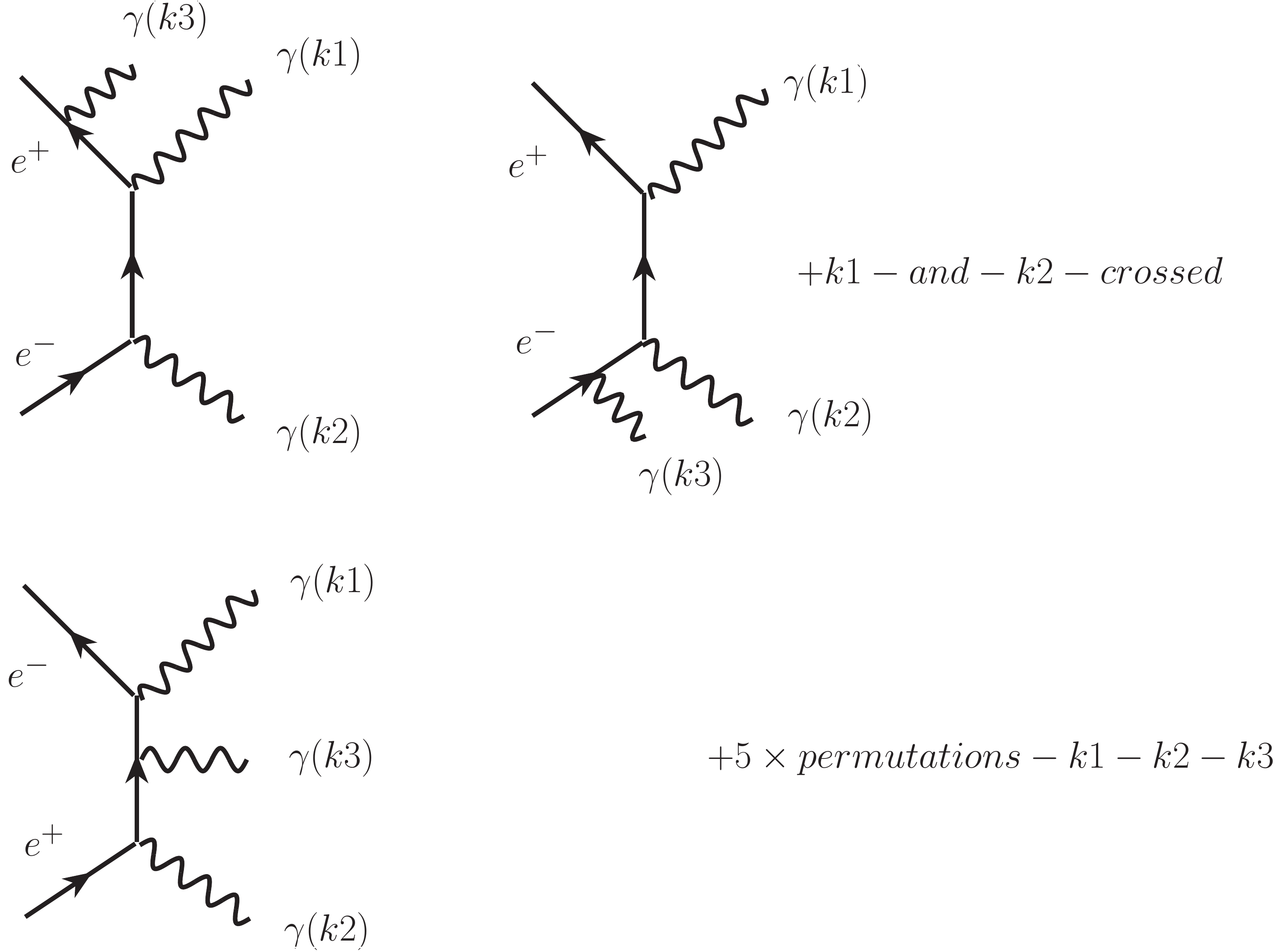}
\end{center}
\caption{ Third order Feynman graphs of the $ \EEGG $  reaction, four soft initial photon corrections and six hard photon corrections. }
\label{CORR2}
\end{figure}

The second set, of four soft real photon initial state corrections, including the six hard photon corrections are 
 shown in the Feynman graphs of Figure~\ref{CORR2} \cite{Fin-graph}.

\subsection{\it The $ \EEGG $ total cross section for hard core and soft core radiative corrections.} 

The differential  cross section for the soft and hard-Bremsstrahlung process is not analytically known. 
The corrections of the Feynman diagrams from Figure~\ref{CORR1} and Figure~\ref{CORR2} are calculated by 
numerical simulations. The details of the differential cross section for the numerical approach are shown in 
appendix .1 and .2 . 

An analytic equation of total integrated cross section, adding the $ \EEG $ plus $ \EEGG $ reactions can be 
calculated (\ref{3Stot1}) and (\ref{3Stot2}). 

\begin{align}
\label{3Stot1}
\sigma _{tot}&=\sigma (2\gamma )+\sigma (3\gamma )\\
\label{3Stot2}
&=\sigma ^{0}+\frac{2\alpha ^3}{s}[\frac{4}{3}v^3-v^2+(\frac{2}{3}\pi ^2-2)v+2-\frac{1}{12}\pi ^2]
\end{align}

The parameter $\nu$ is as function of the mass $ m_{e} $ of the electron and the $ \sqrt{s} $.

\begin{align}
\label{BORNCORR2}
v&=\frac{1}{2}ln\left ( \frac{s}{m^2_{e}} \right )
\end{align}

\subsection{\it The numerical calculation of the $ \EEGG $ cross sections.} 

For practical reasons the differential and total cross section was used from the numerical calculation
for the $ \chi ^2 $ fit. 

\subsubsection{\it The calculation of the $ \EEGG $ differential cross sections.} 

The 6 collaboration under discussion, used an event generator \cite{Berends} for the reaction $ \EEGG $ (\ref{141}).

\begin{equation}
\label{141}
e^+(p_{+})+e^-(p_{-}) \rightarrow \gamma (k_{1})+\gamma (k_{2})+\gamma (k_{3})
\end{equation}

The lowest order differential cross section $ \left ( \frac{d\sigma }{d\Omega } \right )_{Born} $
is corrected numerically by adding the higher order correction of Figure~\ref{CORR1} and Figure~\ref{CORR2} 
to $ O(\alpha ^3) $ in ( \ref{BORNCORR-11} ).

\begin{equation}
\label{BORNCORR-11}
\left ( \frac{d\sigma }{d\Omega } \right )_{\alpha ^3}=\left ( \frac{d\sigma }{d\Omega } \right )_{Born}(1+\delta _{virtual}+\delta _{soft}+\delta _{hard})
\end{equation}

Where, $ \delta _{virtual} $ is the virtual correction and $ \delta_{soft} $ and $ \delta_{hard} $ are the soft- and 
hard-Bremsstrahlung corrections. 

If the energy of the photons from initial state radiation (soft Bremsstrahlung) are too small to be detected, the reaction 
can be treated as 2-photon final state. This is valid if the parameter $ k_{3}/|p_{+}|=k_{0}<<1 $ of the  $ e^+ e^- $ - reaction is 
fulfilled (\ref{142}).

\begin{equation}
\label{142}
e^+(p_{+})+e^-(p_{-}) \rightarrow \gamma (k_{1})+\gamma (k_{2}))
\end{equation}

For the third order differential cross section the program generates  three $ \gamma $ 's events sorted after the energies 
$ E_{\gamma 1 } \geq E_{\gamma 2 } \geq E_{\gamma 3 } $ , with the correct mixture for soft 
$ k_{3}/\vert p_{1}\vert = k_{0} \ll 1 $ and hard QED corrections shown in Figure~\ref{CORR1} and Figure~\ref{CORR2}.
The angle $ \beta $ between the $ E_{\gamma 1 } $ and $ E_{\gamma 2} $ event
 $ \beta_{min} < \beta < \beta_{max} $ is connected to the scattering angle $\theta $ by $ | cos \theta | $.

The differential-cross section  $ \overline{(\frac{d\sigma }{d\Omega })}_{i} $ at 
an angle $ \theta $, an energy $ E_{tot}  $ and  for an angle bin width $ \Delta (|cos\theta |) $
is ( \ref{DIFF1} ).
 
\begin{align}
\label{DIFF1}
\overline{(\frac{d\sigma }{d\Omega })}_{i}=\frac{1}{2\pi \Delta (|cos\theta |)}\sigma _{tot}\frac{N_{i}}{N}
\end{align}

The scattering angle is $ | cos \theta | = ( | cos \theta_{1} | + | \cos \theta_{2} | ) /2 $ , where 
$ \theta_{1} $ is the scattering angle of $ E_{\gamma 1 } $ and  $\theta_{2} $ is the scattering angle of $ E_{\gamma 2 } $,
$ N_{i} $ is the number of events in an angle bin width $ \Delta (|cos\theta |) $ and $ N $ the total amount of
events used in the generator. The Monte Carlo generator together with ( \ref{DIFF1} ) is used to calculate distributions 
of differential cross sections as function of $ | cos \theta | $ including five discussed parameters. 

In the past, the Monte Carlo generator \cite{Berends} was used from all 6 collaborations, to generate the
QED cross section of the he $ \EEGG $ reaction. Our interested is focused on a new Monte Carlo generator BabaYaga@nlo \cite{BABAYAGA}. 
To use the BabaYaga QED generator request a comparison of QED cross section data from both generators.
The new generator implements the same radiative corrections of Figure~\ref{CORR1} and Figure~\ref{CORR2}. 
To generate BabaYaga@nlo $ \EEGG $ events the following seven parameters are used ( \ref{BabaYagaPar} )

\begin{align} 
\label{BabaYagaPar}
p_{11}&=final-\gamma-state=gg\\ \nonumber
p_{22}&=\sqrt{S}-ecms-ENERGY \left [ GeV \right ]\\ \nonumber
p_{33}&=minimum-angle-\theta_{min} - thmin \left [ deg \right ]\\ \nonumber
p_{44}&=maximum-angle-\theta_{max} - thmax \left [ deg \right ]\\ \nonumber
p_{55}&=Accollinear-angle-acoll.-\beta \left [ deg \right ] \\ \nonumber
p_{66}&=Minimum-energy-emin\left [ GeV \right ]  \\ \nonumber
p_{77}&=Number-Generating-Events=N  \nonumber
\end{align}

To perform a $ \chi^{2} $ fit for a finite size of the electron an analytical expression for the differential cross section ( \ref{DIFF1} ) is needed.
The angular distribution of the differential cross section ( \ref{DIFF1} ) is fitted by a $ \chi^{2} $ fit using a polynomial with 6 parameters 
$ p_{1} $  to $ p_{6} $ shown in ( \ref{QEDfit} ).

\begin{align}
\label{QEDfit}
&\left ( \frac{d\sigma }{d\Omega } \right )_{QED}=\left ( \frac{d\sigma }{d\Omega } \right )_{Born} \times \\
&\left ( 1+p_{1}+p_{2}e^{-\frac{x^{1.2}}{2p^2_{3}}}+p_{4}x+p_{5}x^2+p_{6}x^3 \right )|_{x=|cos\theta |} \nonumber
\end{align}

To calculate cross sections for different detectors it is necessary for the calculation of the QED cross sections to
regard the various parameters how every detector is able to measure events. We normalise all cross section
measurements and calculation of the QED cross section to the L3 parameters  ( \ref{BabaYagaL3} ) \cite{L3B}. 

\begin{align} 
\label{BabaYagaL3}
p_{11}&=final-\gamma-state=gg\\ \nonumber
p_{22}&=\sqrt{S}-ecms-ENERGY \left [ GeV \right ]\\ \nonumber
p_{33}&=minimum-angle-\theta_{min} - thmin=16.0 \left [ deg \right ]\\ \nonumber
p_{44}&=maximum-angle-\theta_{max} - thmax=164.0\left [ deg \right ]\\ \nonumber
p_{55}&=Accollinear-angle-acoll.-\beta =90.0\left [ deg \right ] \\ \nonumber
p_{66}&=Minimum-energy-emin=0.48 \left [ GeV \right ]  \\ \nonumber
p_{77}&=Number-Generating-Events=N=100000  \nonumber
\end{align}

Considering the experimental data of the ratio of the total measured cross section divided by the total QED cross
section $ R=\sigma (exp)/\sigma (QED) $ of the L3 collaboration \cite{L3B} Fig. 2 a small deviation from the 
QED - ratio R from 90 GeV to 207 GeV is visible.To test the QED deviation as function of $ \sqrt{s} $ 
we performed two sets of $ \chi^{2} $ tests. One set from 55 GeV to 207 GeV with 17 energy points and one set 
from 91.2 GeV to 200 GeV with 7 energy points.

Including the parameters of ( \ref{BabaYagaL3} ) it is possible to calculate and fit all 17 angular distributions of
the energies under request from $ \sqrt{s} $ = 55 GeV to 207 GeV. The 6 fit parameters  $ p_{1} $  to $ p_{6} $  ( \ref{QEDfit} )
are summarised in Table~\ref{diffpar} ( Upper part ).The same parameters for 7 $ \sqrt{s} $ energies from $ \sqrt{s} $ = 91.2 GeV 
to 200 GeV of the the generator \cite{Berends} are summerized in Table~\ref{diffpar} ( lower part ).

\begin{figure}[htbp]
\vspace{0.0mm}
\begin{center}
 \includegraphics[width=8.5cm,height=8.0cm]{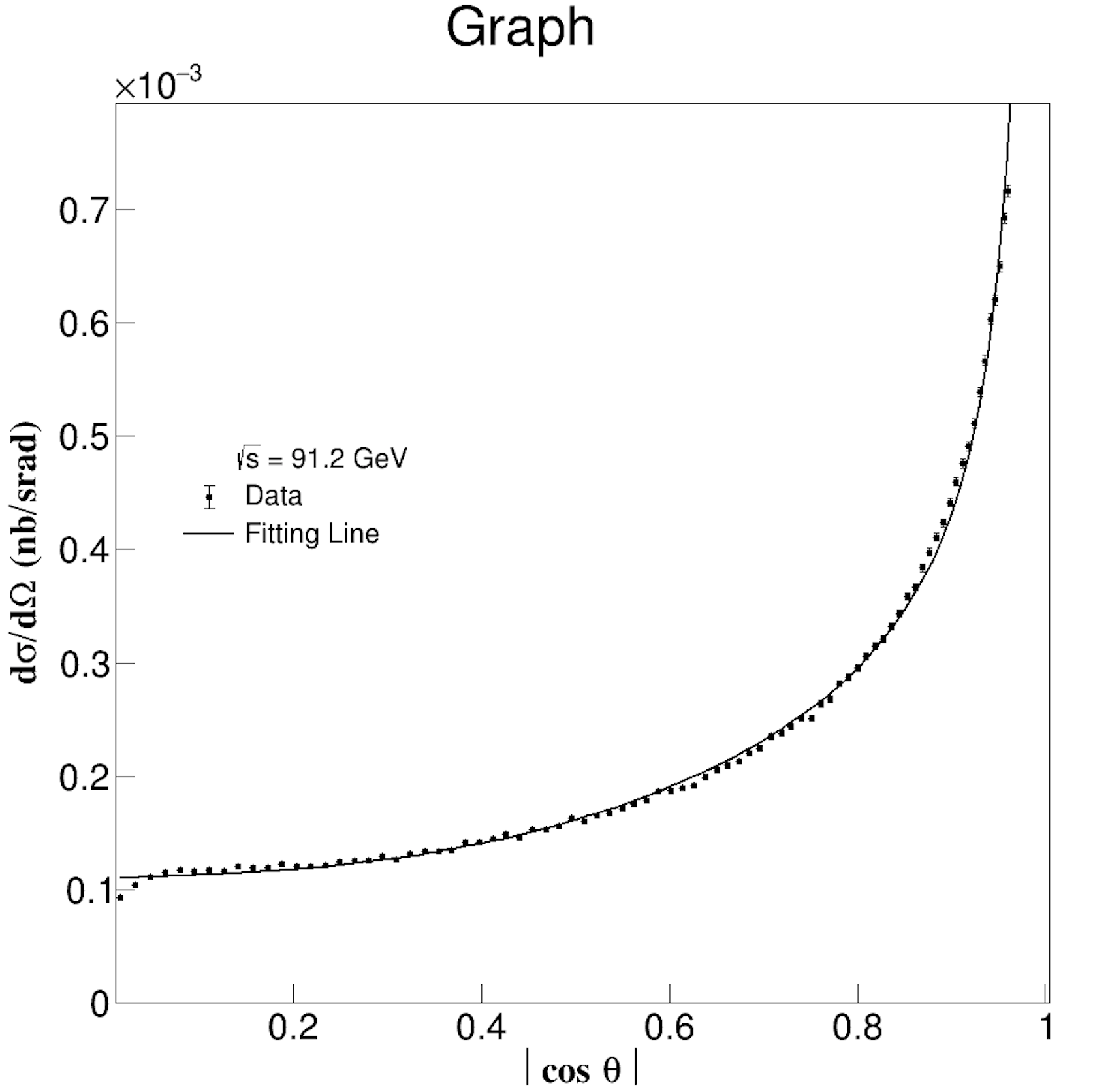}
\end{center}
\vspace{-1.0mm}
\caption{ Differential QED BabaYaga cross section of the $ \EEGG $ reactions, at $ \sqrt{s} = 91.2 GeV $ 
normalised to the parameters of L3. Points are the cross section and black line is the fit. }
\label{QEDdiff}
\end{figure}

 For example at $ \sqrt{s}=91.2 GeV $ the differential QED cross section fitted to ( \ref{QEDfit} ) is shown in 
 Figure ( \ref{QEDdiff} ). 

\begin{figure}[htbp]
\vspace{-10.0mm}
\begin{center}
 \includegraphics[width=9.0cm,height=8.0cm]{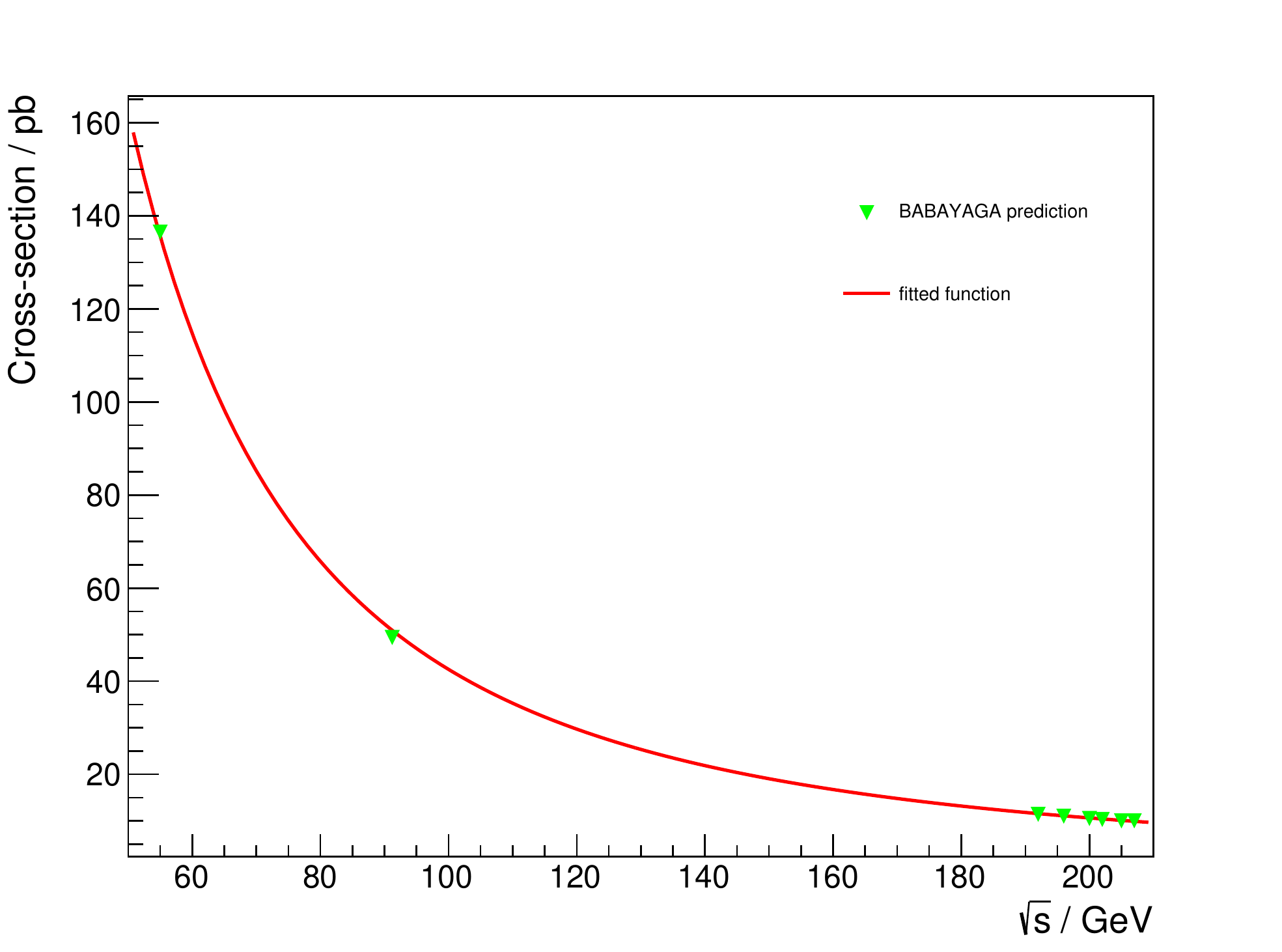}
\end{center}
\caption{ Total QED BabaYaga cross section of the $ \EEGG $ reactions.The points are the cross section and the red line is a fit to this points }
\label{BHABAtot}
\end{figure}

Including the fit parameters from Table~\ref{diffpar} together with (\ref{144}) it is possible to calculate the total
QED BabaYaga cross section from $ \sqrt{s}=55 $ GeV to 207 GeV in Figure ( \ref{BHABAtot} ). The points
are the cross section. 

\begin{equation}
\label{144}
\sigma _{tot}(QED)=\int_{\theta =16.0^{\circ}}^{\theta =164.0^{\circ}}\left ( \frac{d\sigma }{d\Omega } \right )_{QED}d\left | cos\theta  \right |
\end{equation}

The red line is a fit using ( \ref{SIGMAQED11} )and ( \ref{SIGMAQED22} ) to form an analytic expression to the total
QED cross section. The two fit parameters are $ a = - 3.4 \times10^{4} $ and $ b = 4.8\times10^{7} $. 

 
\begin{align}
\label{SIGMAQED11}
\sigma (QED)_{tot} ^{L3}=\sigma(2\gamma )+korr\cdot  \sigma(3\gamma )\\
\label{SIGMAQED22}
korr=a\cdot \sqrt{s}+b
\end{align}

To compare all the experimental data to the QED data the six collaboration used the Monte Carlo QED generator \cite{Berends},
in this analysis we used the generator BabaYaga@nlo \cite{BABAYAGA}. For this reason it was necessary to search for 
deviation between both QED generators. 

The total cross section from the Monte Carlo QED generator \cite{Berends} and the \cite{BABAYAGA} is in the range of the statistical 
error bar the same. At energies between $ \sqrt{s} $ = 91.0 GeV to 207.0 GeV the middle value of \cite{Berends}  is 
$ \sigma _{tot}(QED) $ = 16.2 [pb] and \cite{BABAYAGA}  $ \sigma _{tot}(QED) $ = 16.1 [pb] a total deviation of 
$ \Delta \sigma _{tot}(QED)$ = - 0.1 [pb]. This is approximately the statistical error of the QED generator \cite{Berends}
of $ \Delta \sigma _{tot}(QE D) $ = $\pm$  0.1 [ pb ]. The total cross section $ \sigma _{tot}(QED) $ of \cite{BABAYAGA} above $ \sqrt{s} $ = 91.0 GeV is 
approximately 0.9 $ \% $ smaller as in \cite{Berends}.

\subsection{\it Deviations from QED} 

If the QED is a fundamental theory, the experimental parameters of the $ \EEGG $ reaction should be correctly calculated 
by the Monde Carlo generator \cite{Berends} or \cite{BABAYAGA} up to the Grand Unification scale. All collaboration
\cite{VENUS} to \cite{L33} investigated deviations from the QED $ \EEGG $ reaction from  $ \sqrt{s} $ = 55 GeV up to $ \sqrt{s} $ = 207 GeV.
New non-QED phenomenon could become visible at high energy scales, may be below the  Grand Unification scale. 
An energy scale characterised by a cutoff parameter $ \Lambda $ GeV can be used as a threshold point for a QED breakdown, 
and for underlying new physics. Different models are discussed in \cite{Models1}. Deviations from QED with 
the reaction $ \EEGG $ \cite{ETHZ-USTC}  are investigated in a program, initiated between 1991 and 2020 by the 
Swiss Federal Institute of Technology in Zurich (ETHZ) and USTC Hefei (University of Science and Technology of China). 
In this paper, we focus on a model of finite size of an interaction length in the $ \EEGG $ reaction.

\subsubsection{\it In the direct contact term annihilation a finite interaction length is a measure for the size of the electron} 

The QED differential and total cross section of the $ \EEGG $ reaction is calculated under the condition that
the electron is point-like, without limited interaction length and coupled to the vacuum as shown
in Figure~\ref{CORR1} and Figure~\ref{CORR2}.  So far no experiments exist to test in particular, the non limited interaction length
of the electron up to the Planck scale. If at a certain energy scale $ \Lambda $ between 0 < $ \Lambda $ <  $ M_{P} $
in the $ \EEGG $ reaction, a finite interaction length appears this $ \Lambda $ defines a size of an object the annihilation occurs. 
It is possible to calculate the size of the object via the generalised uncertainty principle \cite{RES:RES1, RES:RES2, RES:RES3} 
or  via the electromagnetic energy E and wave length $ \lambda $ \cite{EMenergy} of the light  the object submits.

It is well known that the wavelength $ \lambda $ of the gammas must be smaller or equal to the size of the interaction
area. If the energy  $ \Lambda $ of the size of the interaction area is known, the frequency $ \nu $ of the gammas is known
via  $ \Lambda=E=\hbar \times \nu $.This frequency $ \nu $  is connected to the size of the object via the wave length $ \lambda $ 
to the equation $ \nu \times \lambda =c $. The energy scale $ \Lambda $ define under these circumstances the size of the 
interaction area and in consequence the size of the electron involved in the annihilation area.

It is possible to construct several effective Lagrangians containing nonstandard $ \gamma e^+ e^- $  or
$ \gamma \gamma e^+ e^- $ couplings which are $ U(1)_{em} $ gauge invariant and only differ in their dimensions \cite{Models1}.
In the lowest order effective Lagrangian, this reaction contains operators of dimension 6, 7 and 8 \cite{Models2}.
In the further discussion, we concentrate on the simplest operator of dimension 6, with the effective Lagrangian of ( \ref{DIR22} ). 

\begin{align}
\label{DIR22}
{\Lagr}_{6}=i\bar{\Psi }\gamma _{\mu }({\vec{D}}_{\nu }\Psi )(g_{6}F^{\mu \nu }+\tilde{g}_{6}\tilde{F}^{\mu \nu })
\end{align} 

The coupling constant $ g_{n}=\sqrt{4\pi }/\Lambda ^{(n-4)} $  $  ( n = 6 ) $, is related to the mass scale $ \Lambda $,
$ D_{\mu }=\partial _{\mu }-ieA_{\mu }$ is the QED covariant derivative, and $ \tilde{F}^{\mu \nu } $ is the dual of the electromagnetic tensor
$ \tilde{F}^{\alpha \beta }=\frac{1}{2}\varepsilon ^{\alpha \beta \mu \nu }F_{\mu \nu } $. The differential cross section to $ {\Lagr}_{6} $
is ( \ref{DIR3} ).

\begin{align}
\label{DIR3}
\left ( \frac{d\sigma }{d\Omega } \right )_{T}&=\left ( \frac{d\sigma }{d\Omega } \right )_{QED}\left [ 1+\delta _{new} \right ] \\
&=\left ( \frac{d\sigma }{d\Omega } \right )_{QED}\left [ 1+\frac{s^2}{2\alpha }\left ( \frac{1}{\Lambda ^4} +
\frac{1}{\tilde{\Lambda ^4}}\right )(1-cos^2\theta ) \right ] \nonumber
\end{align}

We use $ \Lambda =\tilde{\Lambda }=\Lambda _{6} $, higher order terms like $ \Lambda _{7} $ or $ \Lambda _{8} $ of 
$ \delta_{new} $ are omitted. 

To search for a deviation of QED it is common to use a $ \chi^{2} $ tests. This test compares the QED cross section to the experimental
measured cross section. In the test is the QED cross section modified by a non QED direct contact term threshold energy scale $ \Lambda $ after 
equation ( \ref{DIR3} ). 
If the $ \chi^{2} $ test indicates a minimum for a finite threshold energy scale $ \Lambda $ this energy of the cutoff parameter $ \Lambda $  defines via  
the two equations $ \Lambda=E=\hbar \times \nu $ and $ \nu \times \lambda =c $ a finite size of the area the $ e ^ {+} $ $ e ^ {-} $ annihilation must occur. 
For $ \lambda=r_{e} $ is this a measure for size of the electron shown in ( \ref{DIR11} ) including $ \hbar $ the Planck constant and c the speed if light.

\begin{align}
\label{DIR11}
r_{e}=\frac{\hbar c}{\Lambda }
\end{align}

Equation ( \ref{DIR11} ) is generic, the calculation using the generalised uncertainty principle generates the same equation.
It is interesting to notice that in equation ( \ref{DIR11} ) for $ \Lambda \rightarrow \infty $ the size of the object will be $ r_{e}\rightarrow 0 $.
In consequence the point-like QED would be correct to infinite energies.

\section{The measurement of the total cross section $ \sigma _{tot} $ }

\begin{figure}[htbp]
\vspace{-5.0mm}
\begin{center}
 \includegraphics[width=7.0cm,height=7.0cm]{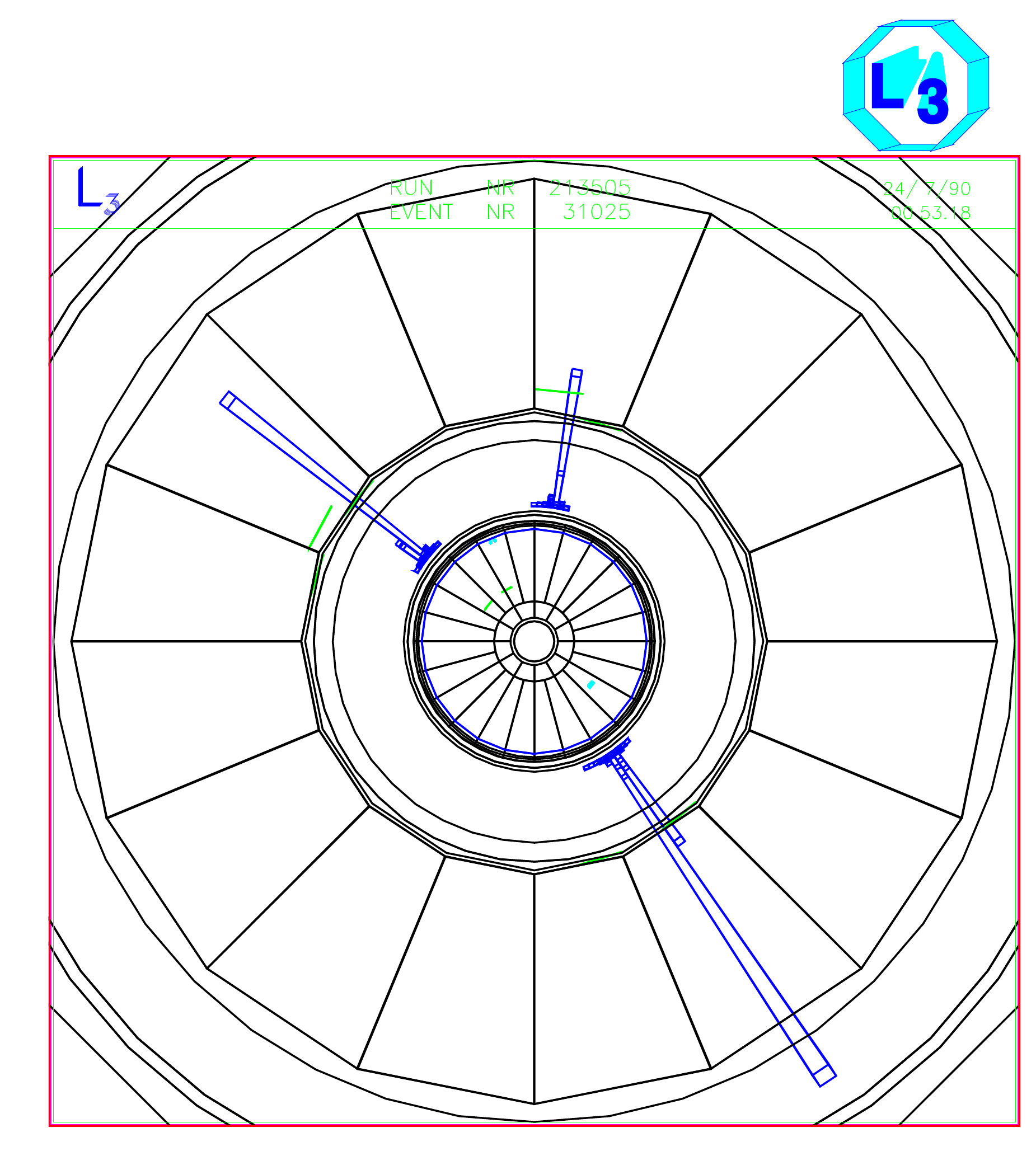}
\end{center}
\caption{ $ \EEGG $ event in the L3 detector at CERN \cite{gamma_L3}}
\label{GAMMA1}
\end{figure}

The $ \EEGG $ reaction initiates in a storage $ e^+ e^- $ ring accelerators a very clean signal in the detector.
For example, Figure~\ref{GAMMA1} shows an event from LEP in the L3 detector. Shown is the position and energy storage of
a  $ \EEGG $ event perpendicular to the $ e^+ e^- $  beam axis in the electromagnetic BGO calorimeter.
The charge sensitive detectors of the inner trackers, the outer hadron calorimeter, and the muon chambers
are free from any signal.


The total cross section is a function of the number of events $ N $ in an angular range of 
the scattering angle  $ 0 < | cos \theta | < | cos \theta |_{max} $, the luminosity $ L $ and the 
efficiency $ \varepsilon $ of the detector for $ \EEG $ events ( \ref{SIGMAtota} ).

\begin{align} 
\label{SIGMAtota}
\sigma _{tot}=\frac{N}{\varepsilon L}
\end{align} 

All seven collaboration measured the total cross section $ \sigma _{tot} $ at different energies $ \sqrt{s} $.
To perform a total $ \chi ^2 $ fit of all experimental data a model is needed to sum over all these information
together in one total cross section at one $ \sqrt{s} $. According to Table~\ref{LUMI} seventeen data sets exist 
from $ \sqrt{s} $ = 55 GeV to 207 GeV. Table~\ref{LUMI} shows the luminosities and the references the total cross section is
is published. At LEP nine data sets exist measuring $ \sigma _{tot} $ from more 
as ONE detector. For eight data sets only ONE detector has measured $ \sigma _{tot} $.  A  total $ \chi ^2 $ 
test would have in total 17 degrees of freedom and if more as one detector is involved in one energy $ \sqrt{s} $ 
the statistical error $ \Delta\sigma(stat) $ would decrease.

All the detectors of the six collaboration use different scattering angles  $ 0 < | cos \theta | < | cos \theta |_{max} $, 
luminosities $ L $ and the efficiencies $ \varepsilon $ for the for $ \EEG $ events. In common all the collaborations used for 
 $ \chi ^2 $ tests the CERNLIB program MINUIT  \cite{MINUIT} or Monte Carlo programs using the same radiative 
 correction as discussed in Figure~\ref{CORR1} and Figure~\ref{CORR2}. The Monte Carlo program generates event
 number including different scattering angles  range $ 0 < | cos \theta | < | cos \theta |_{max} $  and efficiency $ \varepsilon $.
 This fact allows to normalise the experimental data of the six collaborations to an imaginary L3 detector using the 
 QED cross section of the different collaborations.
 
 \subsection{\it Calculation of $ \sigma (tot) $, $ \Delta \sigma (stat) $, ratio R(exp) and $ \Delta R(stat) $ for more than one detector.}  

At LEP nine data sets exist measuring $ \sigma _{tot} $ from more as ONE detector. To calculate 
the total experimental cross section $ \sigma(exp)_{tot}^{(sum)} $ ( \ref{SIGMAtot4} ) , the statistical error 
$ \Delta\sigma(exp)_{tot}^{(sum)} $ (\ref{SIGMAtot5}) , the ratio  R(exp) = $ R^{sum } $  ( \ref{SIGMAtot6} ) and
the statistical error of this ratio $ \Delta R(stat) $ = $ \Delta R(stat)^{(sum)} $ ( \ref{SIGMAtot8} ) the following 
INPUT parameters are needed. 

The total experimental cross section $ \sigma(exp)_{tot}^{(det_i)} $ [pb] from VENUS, TOPAS, ALEPH, DELPHI, OPAL 
and total L3-N(exp) event rate from Table~\ref{EXPtot}. The total QED cross section $ \sigma(QED)_{tot}^{(det_i)} $ [pb] 
from VENUS, TOPAS, ALEPH, DELPHI, OPAL and total L3-$ N(QED)^{L3} $ event rate from Table~\ref{QEDtot} and
the luminosity used from the VENUS, TOPAS, ALEPH, DELPHI, L3 and OPAL experiment from Table~\ref{LUMI} .

Under this condition in a first step, the model normalise the experimental total cross section of a detector$_{i} $ $ \sigma(exp)_{tot}^{(det_i)} $ 
at one energy $ \sqrt{s} $ to the L3 normalised experiential total cross section  $ \sigma(exp)_{tot}^{(L3-det_i)} $ ( \ref{SIGMAtot1} ) 
via the total QED cross section of L3 $ \sigma(QED)_{tot}^{(L3)} $ to the total QED cross section of the 
detector detector$_{i} $ $ \sigma(QED)_{tot}^{(det_i)} $ at $ \sqrt{s} $ under investigation.
For the detail calculation are the numerical values of $ \sigma(exp)_{tot}^{(det_i)} $ summarised, in Table~\ref{EXPtot}
including the statistical error $ \Delta\sigma (exp)_{tot}^{det} $ are needed.
 
The values of $ \sigma(QED)_{tot}^{(det_i)} $ are summarised in Table~\ref{QEDtot}. The
value $ \sigma(QED)_{tot}^{(L3)} $ is calculated via ( \ref{SIGMAQED11} ) , ( \ref{SIGMAQED22} ) and for
simplicity also added in Table~\ref{QEDtot}.

Secondly the model introduces a to - L3 QED normalised efficiency $ \varepsilon^{L3} $, including the L3 QED - event numbers
$ N(QED)^{L3} $ of the total cross section at the  $ \sqrt{s} $ - energy under investigation, the total QED cross section
 $\sigma(QED)_{tot}^{(L3)}$ and the L3 luminosity $ L^{ (L3) }$ at $ \sqrt{s} $ ( \ref{SIGMAtot2} ). The numerical values
 of $ N(QED)^{L3} $ are taken from \cite{L3B}. Inserted from this reference are the numbers from page 33 Table 2 : " The 
 expected 2$\gamma$ events ", which is in agreement with QED - Monde Carlo generators. The 3$\gamma$ is
 in the generator already implicit included in the parameter $ p_{11}=final-\gamma-state=gg $ and 
 $ p_{55}=Accollinear-angle-acoll.-\beta =90.0\left [ deg \right ] $. The values for L3 luminosity $ L^{ (L3) }$ 
 at the $ \sqrt{s} $ energy under investigation are taken from Table~\ref{LUMI}.
 
Third, the to - L3 normalised total QED cross section of the detector detector$_{i} $ $ \sigma(QED)_{tot}^{(det_i)} $ at $ \sqrt{s} $ 
under investigation ( \ref{SIGMAtot1} ), the L3 QED efficiency $ \varepsilon^{L3} $ ( \ref{SIGMAtot2} )
and the luminosity of the different detectors $ L^{(det)}_{i} $ open the possibility to calculate the experimental 
counting rate of every detector normalised to L3 of this detector $ N(exp)_{i}^{L3} $ ( \ref{SIGMAtot3} ).
The numerical values for $ L^{(det)}_{i} $ , are summarised in Table~\ref{LUMI}. In this Table included
are the important references for Table~\ref{EXPtot} and Table~\ref{QEDtot}.

Including the detailed numerical numbers from (\ref{SIGMAtot1}), (\ref{SIGMAtot2}) and ( \ref{SIGMAtot3} )
it is possible to sum over $ N(exp)_{i}^{L3} $ and $ L^{(det)}_{i} $ to calculate for every $ \sqrt{s} $ under investigation
the total summed cross section $ \sigma(exp)_{tot}^{(sum)} $ ( \ref{SIGMAtot4} ) 
and the statistical error 
\newline
$ \Delta\sigma(exp)_{tot}^{(sum)} $ ( \ref{SIGMAtot5} ).

The total summed cross section $ \sigma(exp)_{tot}^{(sum)} $ ( \ref{SIGMAtot4} ) divided by the 
total QED cross section of L3 $ \sigma(QED)_{tot}^{(L3)} $ allowes to calculate the very important
ratio $ R^{sum} $ at the $ \sqrt{s} $ under investigation ( \ref{SIGMAtot6} ).

To calculate the statistical error $ \Delta R(stat)^{(sum)} $ ( \ref{SIGMAtot8} ) the most conservative approach
via the maximal possible error $ \Delta Max R $ is used  ( \ref{SIGMAtot7} ).
  
\begin{align}
\label{SIGMAtot1}
\sigma(exp)_{tot}^{(L3-det_i)} =\sigma(exp)_{tot}^{(det_i)}\frac{\sigma(QED)_{tot}^{(L3)}}{\sigma(QED)_{tot}^{(det_i)}}\\
\label{SIGMAtot2}
\varepsilon^{L3}=\frac{N(QED)^{L3}}{(\sigma(QED)_{tot}^{(L3)}L^{(L3)})}\\ 
\label{SIGMAtot3}
N(exp)_{i}^{L3}= \sigma(exp)_{tot}^{(L3-det_{i})}\varepsilon^{L3}L^{(det)}_{i} \\
\label{SIGMAtot4}
\sigma(exp)_{tot}^{(sum)} =\frac{\sum_{i}N(exp)_{i}^{L3}}{[\varepsilon^{L3}\sum_{i}L_{i}^{det}]}\\
\label{SIGMAtot5}
\Delta\sigma(exp)_{tot}^{(sum)} =\frac{\sqrt{\sum_{i}N(exp)_{i}^{L3}}}{[\varepsilon^{L3}\sum_{i}L_{i}^{det}]}\\
\label{SIGMAtot6}
R^{sum}=\frac{\sigma(exp)_{tot}^{(sum)}}{\sigma(QED)_{tot}^{(L3)}}\\
\label{SIGMAtot7}
\Delta Max R = \frac{\sigma (exp)_{_{tot}}^{sum}+\Delta \sigma(exp)_{tot}^{sum}}{\sigma(QED)_{tot}^{L3} }\\
\label{SIGMAtot8}
\Delta R(stat)^{(sum)}=\Delta Max R -R^{sum}
\end{align}
  
Not all different collaborations mark for $ \sigma(QED)_{tot}^{(det_i)} $ a systematic error, for this reason the 
( \ref{SIGMAtot5} ) and ( \ref{SIGMAtot8} ) include only the statistical error, which is dominating in this analysis.  
All collaborations generate with a Monte Carlo generator many millions of events to calculate the differential or 
total QED $e^{+}e^{-}\to\gamma\gamma$ cross sections. This allows us to keep the systematic error originated from MC 
negligible, compared to the data statistical error. 
This analysis investigated , one more possible important systematic error. We investigated two different Monte Carlo 
generators and used different energy ranges for the $\chi^2$ test to study the systematic error. In the following 
sections, we will see that the interaction radius from all the four different tests, are in the range of the 
statistical error the same ( Table~\ref{Comparison} ). Interesting is that even the completely independent analyses 
of the different cross section agrees with this interaction radius. ( Table~\ref{Comparison} ). It proves that the 
systematic error is not changing the result of the analysis.

  \subsection{\it Calculation of $ \sigma (tot) $, $ \Delta \sigma (stat) $, ratio R(exp) and $ \Delta R(stat) $ for one detector.} 
  
  The INPUT data in the case ONLY one detector contributes to the calculation of ( \ref{SIGMAtot9} ) to ( \ref{SIGMAtot13} )
  the detail information is again included in Table~\ref{EXPtot}, Table~\ref{QEDtot} and Table~\ref{LUMI}. 
  In this case eight data sets exist of $ \sigma _{tot} $ .
  
  Under these conditions the total experimental cross section normalised to L3 $ \sigma (exp)_{tot}^{(single)} $ ( \ref{SIGMAtot9} ) 
  is like $ ( \ref{SIGMAtot1} ) $ a function of the total experimental cross section of one detector $ \sigma (exp)_{tot}^{det} $ and the ratio of the total QED L3 
  cross section $ \sigma (QED)_{tot}^{L3} $ to the total QED detector cross section $ \sigma (QED)_{tot}^{det} $. It is similar 
  possible to calculate from the experimental error of detector $ \Delta\sigma (exp)_{tot}^{det} $ via the ratio $ \sigma (QED)_{tot}^{L3} $ to 
  $ \sigma (QED)_{tot}^{det} $ the to L3 normalised error  $ \Delta\sigma (exp)_{tot}^{(single)} $ ( \ref{SIGMAtot10} ). 
  The ratio R $^{single} $ ( \ref{SIGMAtot11} ) is 
  a function of ( \ref{SIGMAtot9} ) and the total QED L3 cross section $ \sigma (QED)_{tot}^{L3} $. The $ \sigma (QED)_{tot}^{L3} $ 
  cancels in ( \ref{SIGMAtot11} ), what replaces $ \sigma (exp)_{tot}^{(single)} $ by $ \sigma (exp)_{tot}^{det} $ and
  $ \sigma (QED)_{tot}^{L3} $  to $ \sigma (QED)_{tot}^{det} $.
  To calculate the statistical error $ \Delta R(stat)^{(single)} $ ( \ref{SIGMAtot13} ) for simplicity the maximum value of 
  $ \Delta Max R^{(single)} $ ( \ref{SIGMAtot12} ) is used to form the difference between $ \Delta Max R^{(single)} $
  and $ R^{single} $. In ( \ref{SIGMAtot12} ) is the sum of ( \ref{SIGMAtot9} ) and ( \ref{SIGMAtot10} ) devided by $ \sigma (QED)_{tot}^{L3} $.
    
\begin{align}
\label{SIGMAtot9}
\sigma (exp)_{tot}^{(single)}=\sigma (exp)_{tot}^{det}\frac{\sigma (QED)_{tot}^{L3}}{\sigma (QED)_{tot}^{det}}\\
\label{SIGMAtot10}
\Delta\sigma (exp)_{tot}^{(single)}=\Delta\sigma (exp)_{tot}^{det}\frac{\sigma (QED)_{tot}^{L3}}{\sigma (QED)_{tot}^{det}}\\
\label{SIGMAtot11}
R^{single}=\frac{\sigma (exp)_{tot}^{single}}{\sigma (QED)_{tot}^{L3}} =\frac{\sigma (exp)_{tot}^{det}}{\sigma (QED)_{tot}^{det}}\\
\label{SIGMAtot12}
\Delta Max R^{(single)} = \frac{\sigma (exp)_{_{tot}}^{single}+\Delta \sigma(exp)_{tot}^{single}}{\sigma(QED)_{tot}^{L3} }\\
\label{SIGMAtot13}
\Delta R(stat)^{(single)}=\Delta Max R^{single} -R^{single}
\end{align}

Not all different collaborations mark for $ \sigma(QED)_{tot}^{(det)} $ a systematic error, for this reason the 
( \ref{SIGMAtot10} ) and ( \ref{SIGMAtot12} ) include only the statistical error.

 \subsection{\it Numerical calculation of $ \sigma (tot) $, $ \Delta \sigma (stat) $, ratio R(exp) and $ \Delta R(stat) $.} 

Inserting the numerical values from Table~\ref{EXPtot}, Table~\ref{QEDtot} and Table~\ref{LUMI} in 
( \ref{SIGMAtot1} ) until  ( \ref{SIGMAtot13} ) allows to calculate $ \sigma (tot) $, $ \Delta \sigma (stat) $, ratio R(exp) and $ \Delta R(stat) $
shown in  Table~\ref{table6}.
    
\begin{table}
\begin{center}
\caption{ Summary $ \sigma (tot) $, $ \Delta \sigma (stat) $ and ratio R(exp) and $ \Delta R(stat) $} 
\label{table6}
\begin{tabular}{ | l | l | l | }
  \hline
  $ \sqrt{s} $ [ GeV ] &\multicolumn{1}{ | c | }{ $ \sigma (tot) $ $ \Delta \sigma (stat) $ [ pb ] }&\multicolumn{1}{ | c | }{ R(exp) $ \Delta R(stat) $  }\\
  \hline
   55   & 124.7   $ \pm  $ 13.2    & 0.92  $ \pm $ 0.10 \\ \hline
   56   & 150.6   $ \pm  $ 9.7     & 1.15  $ \pm $ 0.07 \\ \hline
   56.5 & 141.6   $ \pm  $ 22.9    & 1.10  $ \pm $ 0.18 \\ \hline
   57   & 135.5   $ \pm  $ 10.8    & 1.07  $ \pm $ 0.09 \\ \hline
   57.6 & 125.3   $ \pm  $ 2.0     & 1.01  $ \pm $ 0.02 \\ \hline
   91   & 50.3    $ \pm  $ 0.9     & 0.99  $ \pm $ 0.02 \\ \hline
   133  & 26.5    $ \pm  $ 5.8     & 1.10  $ \pm $ 0.24 \\ \hline
   162  & 16.1    $ \pm  $ 2.4     & 0.98  $ \pm $ 0.15 \\ \hline
   172  & 15.6    $ \pm  $ 2.6     & 1.08  $ \pm $ 0.18 \\ \hline
   183  & 12.6    $ \pm  $ 0.3     & 0.99  $ \pm $ 0.03 \\ \hline
   189  & 11.8    $ \pm  $ 0.2     & 0.99  $ \pm $ 0.02 \\ \hline
   192  & 11.0    $ \pm  $ 0.5     & 0.95  $ \pm $ 0.04 \\ \hline
   196  & 11.3    $ \pm  $ 0.3     & 1.02  $ \pm $ 0.03 \\ \hline
   200  & 10.1    $ \pm  $ 0.3     & 0.95  $ \pm $ 0.03 \\ \hline
   202  & 10.1    $ \pm  $ 0.4     & 0.97  $ \pm $ 0.04 \\ \hline
   205  & 10.0    $ \pm  $ 0.3     & 0.99  $ \pm $ 0.03 \\ \hline
   207  & 9.7     $ \pm  $ 0.2     & 0.98  $ \pm $ 0.02 \\ \hline
 \end{tabular}
\end{center}
\end{table}
     
\begin{figure}[htbp]
\vspace{0.0mm}
\begin{center}
 \includegraphics[width=8.5cm,height=6.0cm]{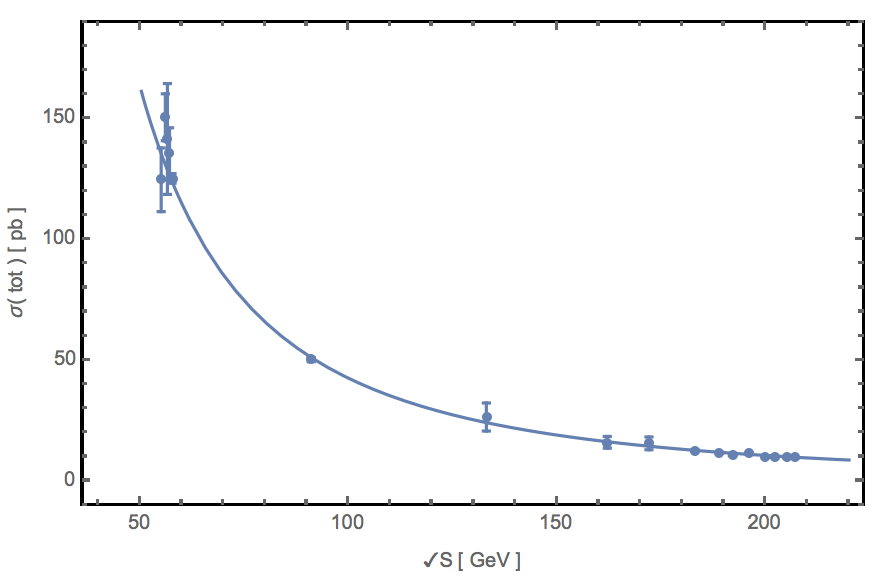}
\end{center}
\caption{  The $ \sigma $(tot)  of the $ \EEGG $ reaction of all detectors as function of centre-of-mass energy $ \sqrt{s} $. The data (points) are compared to QED prediction (solid line).}
\label{SIGMAtot}
\end{figure}

The  $ \sigma (tot) $ values from $ \sqrt{s} $ = 55 GeV to 207 GeV together with  $ \Delta \sigma (stat) $ of Table~\ref{table6} 
compared to the total QED cross section $  \sigma(QED)_{tot}^{(L3)} $  ( \ref{SIGMAQED11} ) is displayed 
in Figure~\ref{SIGMAtot}. 
 

 Figure~\ref{SIGMAtot} shows, in the range of the sensitivity between the experimental measured values $ \sigma(tot) $ including the
 statistical error $ \Delta \sigma (stat) $ and the total QED cross section 
 \newline
 $ \sigma(QED)_{tot}^{(L3)} $ a good agreement. 
 
\begin{figure}[htbp]
\vspace{0.0mm}
\begin{center}
 \includegraphics[width=8.5cm,height=6.0cm]{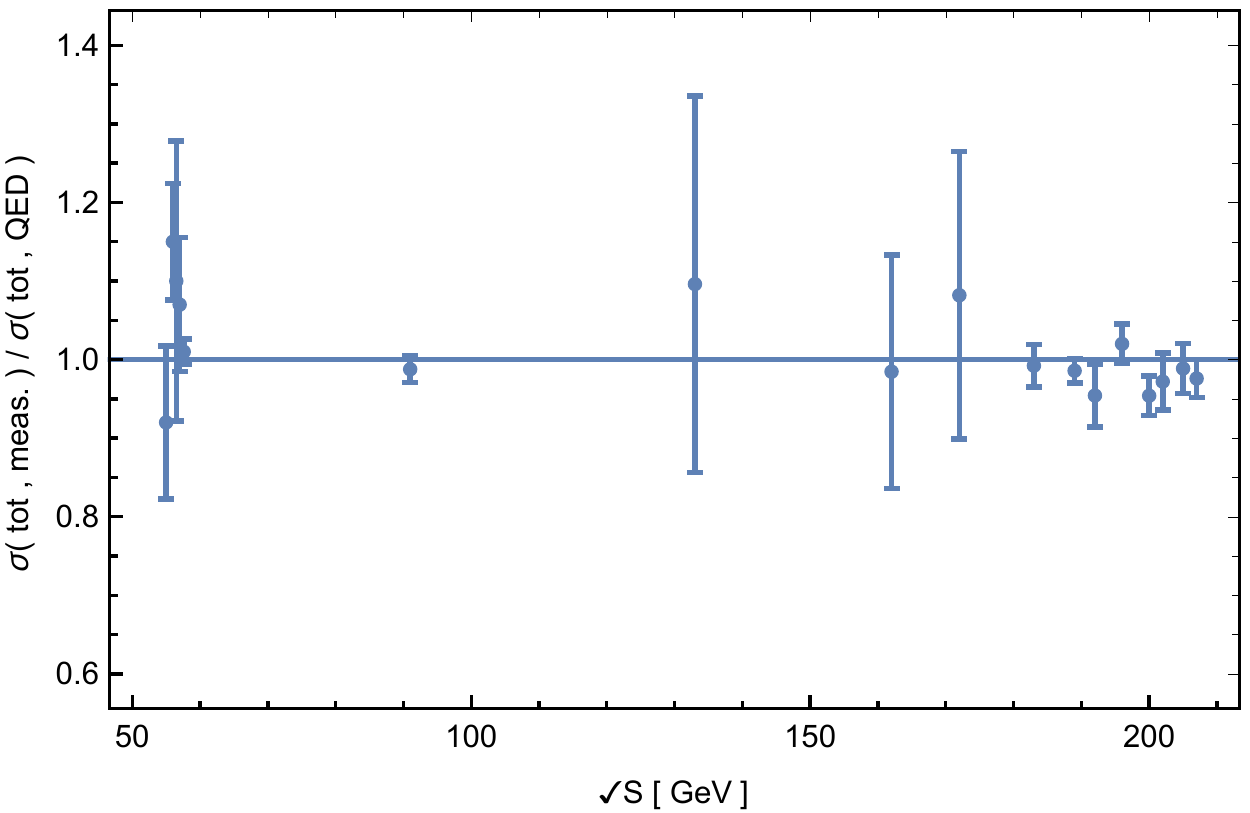}
\end{center}
\caption{  Ratio $ \sigma $ (tot,meas.) /  $ \sigma$(tot,QED)  of the $ \EEGG $ reaction of all detectors as function of  centre-of-mass energy $ \sqrt{s} $ .
                The experimental data (points) and QED prediction (solid blue line).}
\label{Ratio-1}
\end{figure}

To search for deviation between measured values $ \sigma(tot) $ and the total QED cross section $ \sigma(QED)_{tot}^{(L3)} $ the graphic
of Figure~\ref{SIGMAtot} is not sensitive enough because the deviation is on the $ \% $ level. A more sensitive graphic Figure~\ref{Ratio-1} is to display the 
ratio $ \sigma $ (tot,meas.) /  $ \sigma$(tot,QED)  of the $ \EEGG $ reaction of all detectors as function of  centre-of-mass energy $ \sqrt{s} $.
At $ \sqrt{s} $ > 90 GeV a negative interference between measure values and QED of some $ \% $ is visible. 

In addition is visible that the statistical error for 8 data point is bigger as for 9 data points. 
The decrease of the statistical error originated from the fact that, at LEP nine data sets exist measuring $ \sigma _{tot} $ 
from more as ONE detector ( \ref{SIGMAtot5} ) and eight data sets exist with only ONE detector has measured $ \sigma _{tot} $.
For example,  is in the used model at $ \sqrt{s} $ = 207 GeV the statistical error $ \Delta \sigma (stat) $ = 0.2 [ pb ],
compared to the statistical error of L3 $ \Delta \sigma (stat) $ = 0.34 [ pb ]  at the same $ \sqrt{s} $ = 207 GeV  \cite{L3B}. 
This is a decrease of approximately 42 $ \% $.
  
 
\section{ Search for finite size of electron using a $ \chi^{2} $ test  of the ratio $ \sigma $ (tot,meas.) /  $ \sigma$(tot,QED) . }

The signal for the finite size of the electron is weak, not visible in the graphic of the total experimental cross section compared with the QED
total cross section in Figure~\ref{SIGMAtot}. A deviation of approximately some $ \% $ is visible in the graphic of the ratio 
$ \sigma $ (tot,meas.) /  $ \sigma$(tot,QED)  of the $ \EEGG $ reaction in Figure~\ref{Ratio-1}. In accordance with the sensitivity between 
experiment and QED, a $ \chi^{2} $ test on this ratio ( \ref{CHI-1} ) is performed to search for a minimum in $ \chi^{2} $.

\begin{align}
\label{CHI-1}
\chi ^{2}&=\sum_{i}\left \{ \frac{R(E_{i};exp)-R(E_{i};QED;\Lambda )}{\Delta R(E_{i};exp)} \right \}^{2}
\end{align} 

The ratio of the experimental data $ R(E_{i};exp) $ is in accordance with Table~\ref{table6}, the ratio of $ \sigma $ (tot,meas.) /  $ \sigma$(tot,QED) 
at a $ \sqrt{s} $ energy $ E_{i} $. The statistical error $ \Delta R(E_{i};exp) $ is the error at $ \sqrt{s} $ energy $ E_{i} $, in accordance with 
Table~\ref{table6}. The term to search for a deviation of a finite size of the electron is  $ R(E_{i};QED;\Lambda) $ ( \ref{Ratio-11} ). 
Included is the fit parameter  $ \left ( 1/\Lambda _{6} \right )^{4} $ , a function to the interaction size of the electron $ r_{e} $ ( \ref{DIR11} ).

\begin{align}
\label{Ratio-11}
R(E_{i};QED;\Lambda )=\frac{\int_{\Omega _{min}}^{\Omega _{max}}\left ( \frac{d\sigma }{d\Omega } 
\right )_{(E_{i};QED;\Lambda )}d\Omega }{\int_{\Omega _{min}}^{\Omega _{max}}\left ( \frac{d\sigma }{d\Omega } \right )_{(E_{i};QED)}d\Omega }
\end{align}

Essential for the whole program of the $ \chi^{2} $ test is to use the fit parameter  $ \left ( 1/\Lambda _{6} \right )^{4} $. This parameter is sensitive 
to the theoretical calculation for a deviation from the QED differential cross section for positive and negative interference. A parameter $ \Lambda _{6} $ 
would test ONLY the positive interference and cut out the negative part. 

These problem is visible in ( \ref{Ratio-11} ) and in particular ( \ref{NomRatio-11} ). The parameter $ \Lambda _{6} $ would keep the sign in front of 
$ (1-cos^{2}\theta ) $ always positive independent from the $ \pm $ sign from $ \Lambda _{6} $. As consequence would be always
$ \left ( d\sigma /d\Omega  \right )_{_{(E_{i};QED;\Lambda )}}>\left ( d\sigma _{}/d\Omega _{(E_{i};QED)} \right ) $ and
$ R(E_{i};QED;\Lambda )>1 $ . The $ \chi^{2} $ test would be NOT able to find values $ R(E_{i};QED;\Lambda )<1 $.
For this reason a negative interference could NOT be detected.

\begin{align}
\label{NomRatio-11}
&\int_{\Omega _{min}}^{\Omega _{max}}\left ( \frac{d\sigma }{d\Omega } \right )_{(E_{i};QED;\Lambda) }d\Omega =\\
&\int_{\Omega _{min}}^{\Omega _{max}} \left [ \left ( \frac{d\sigma }{d\Omega }\right )_{(E_{i};QED)}(1+\frac{s^{2}}
{\alpha \Lambda _{6}^{4}}(1-cos^{2}\theta )) \right ]d\Omega \nonumber
\end{align}

The integrals ( \ref{Ratio-11} and  \ref{NomRatio-11} ) included the differential QED cross section $ \left ( d\sigma /d\Omega  \right )_{(E_{i};QED)} $. 

To test the angular contribution part of $ R(E_{i};QED;\Lambda ) $ function it is possible to integrate ( \ref{Ratio-11} ) over the  $ \sqrt{s} $ and
use only the contribution of the angular distribution of the direct contact term ( \ref{DIR3} ) like ( \ref{Ratio-k} ).

\begin{align}
\label{Ratio-k}
R(E_{i};QED;\Lambda )=\\ \nonumber
\frac{\int_{\Omega _{min}}^{\Omega _{max}}\left ( \frac{d\sigma }{d\Omega } \right )_{(E_{i};QED)}\left [ 1+\frac{s^{2}}{\alpha \Lambda _{6}^{4}}
(1-cos^{2}\theta ) \right ]d\Omega}{\int_{\Omega _{min}}^{\Omega ^{max}}\left ( \frac{d\sigma }{d\Omega } \right )_{(E_{i;QED})}d\Omega }=\\
\frac{k_{1}\int_{\Omega _{min}}^{\Omega ^{max}}\left [ 1+\frac{s^{2}}{\alpha \Lambda _{6}^{4}}(1-cos^{2}\theta ) \right ]d\Omega }
{k_{2}\int_{\Omega _{min}}^{\Omega ^{max}}d\Omega } \nonumber
\end{align}

The energy contribution of $ R(E_{i};QED;\Lambda ) $ is in the range of $ \% $ level, what opens the possibility to investigate the 
$ \chi^{2} $ test under the assumption that the constant $ k_{1} $ is approximately $ k_{2} $.

\subsection{ Numerical calculation of the  $ \chi^{2} $ tests }

To investigate the sensitivity of the $ \chi^{2} $ test to the Monte Carlo generator BabaYaga@nlo \cite{BABAYAGA} and the generator \cite{Berends} a separate 
numerical calculation for both test was performed. In addition, an approximation was used to test only the impact of the direct contact term ( \ref{DIR3} )
to the  $ \chi^{2} $ test .

The mathematical details of different possible calculation of significance and error of the interaction radius $ r $ is studied in 
Appendix 4 .

\subsubsection{ Calculation of the $ \chi^{2} $ test with QED BabaYaga }

\begin{figure}[htbp]
\vspace{-3.0mm}
\begin{center}
 \includegraphics[width=8.0cm,height=7.0cm]{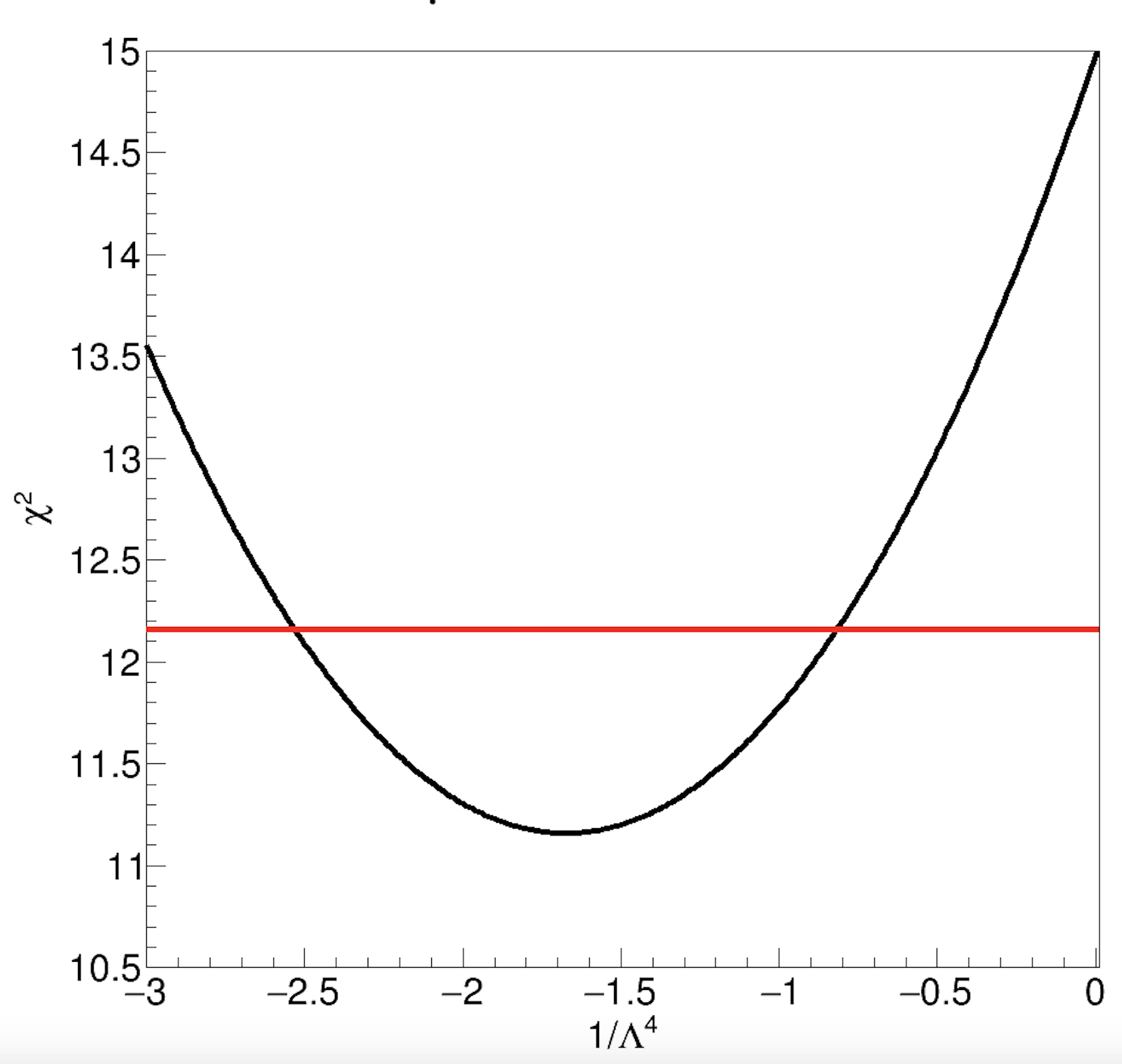}
\end{center}
\vspace{-2.0mm}
\caption{  $ \chi^{2} $ test on the ratio ( \ref{CHI-1} ) of the $ \EEGG $ reaction from centre-of-mass energy $ \sqrt{s} $ = 55 GeV to 207 GeV using
               QED data from Monte Carlo generator BabaYaga@nlo . The red line shows  $ \sigma $ limit. }
\label{CHIBhabha}
\end{figure}

The generators under discussion generate numbers of events respecting to L3 parameters  ( \ref{BabaYagaL3} ).
The events per angular range are used to fit a differential cross section as function of 
$ d\sigma /d\Omega \left [ nb/srad \right ]$ 
and $ \left | cos \theta  \right | $ . An example of such a differential cross section at $ \sqrt{s} $ = 90.2 GeV is shown in Figure (\ref{QEDdiff}).
The numerical parameters of this fit $ p_{1} $  to $ p_{6} $ defined in ( \ref{QEDfit} ) for the 17 $ \sqrt{s} $ energies from 55 GeV to 207 GeV
are summarised in Table~\ref{diffpar} ( Upper part ). The $ \chi^{2} $ test on the ratio ( \ref{CHI-1} ) as function of  $ \chi^{2} $ and $ 1/\Lambda ^{4} $ 
of the $ \EEGG $ reaction from centre-of-mass energy $ \sqrt{s} $ = 55 GeV to 207 GeV using QED data from Monte Carlo generator BabaYaga@nlo 
is displayed in Figure (\ref{CHIBhabha}).

The minimum of the $ \chi^{2} $ test including the error of Figure (\ref{CHIBhabha}) is $ 1/\Lambda ^{4} =-1.62_{-0.83}^{+0.84}\times 10^{-13}GeV^{-4} $.
The $ \chi^{2} $ value at the minimum is $ \chi^{2} $ = 11.21. The fit uses 17 degrees of freedom 
according to Table~\ref{table6}.

The important result of the $ \chi^{2} $ fit is that the fit parameter $ 1/\Lambda ^{4} $ has a negative sign. According to the 
Figure~\ref{Ratio-1} is the fit sensitive, to the fact that the total QED - cross section is bigger as the experimental total cross section
above approximately $ \sqrt{s} $ = 92 GeV to 207 GeV. The discussed direct contact interaction term $ \delta _{new} $ ( \ref{DIR3} ) has a negative
sign. This indicates a negative interference of the direct contact interaction in the  $ {\EEG} $ reaction.

Essential of the $ \chi^{2} $ fit result is in particular the significance $ \sigma $ of the fit. After international rules the physic community 
accepts a $ \sigma > 5 $ as a discovery of new physics and a $ \sigma < 5 $ as a hint of new physics. 

A detailed mathematical calculation of significance is discussed in Appendix 4.1 " Equations for the calculation of the significance of a $ \chi^{2} $ test. ".
A first direct approximation of the significance of the $ \chi^{2} $ fit can be direct estimated from the error bars of the fit
$ \sigma =A/\Delta A $  = 1.62 / 0.83 = 1.95. In a second approximation to calculate the significance $ \sigma $ a statistical probability 
function $ p $ ( \ref{pvalue1} ) is used. The $ p $ value is in the discussed $ \chi^{2} $ for 17 degrees of freedom and the minimum 
is $ \chi^{2} $ = 11.2 equal  $ p $ = 0.15. According to Figure (\ref{pvaluesmall}), the significance is approximate $ \sigma $ = 1.5.

Similar to the significance, a detailed mathematical calculation of the error $ \Delta r_{e} $ of the interaction size $ r_{e} $ of the electron in the  
$ \chi^{2} $ test is discussed in the Appendix 4.2 " The error $ \Delta r_{e} $ of the interaction size $ r_{e} $ of the electron in the  $ \chi^{2} $ test. ".
According to ( \ref{DIR11} ), the size of the interaction term is $ r_{e} =1.25 \times 10^{-17} $ [ cm ], and 
the error $ \Delta r_{e} = 0.16 \times 10^{-17}$  [ cm ]. The summary of all these results is given in of Table~\ref{Baba}.

 \begin{table}
 \begin{center}
 \caption{ Summary of $ \chi^{2} $ test with BabaYaga@nlo generator. }
 \label{Baba}
 \begin{tabular}{ | l | c | r | r| }
 \hline
                    & $ (1/\Lambda )^{4} [GeV^{-4}] $     & $\sigma$ & $ ( r \pm \Delta r ) \times 10^{-17} [cm] $  \\  \hline
direct           & $ (-1.62 \pm 0.83)\times 10^{-13} $ & 1.95 &  \\  \hline
p - value      & $ (-1.62 \pm 0.83)\times 10^{-13} $ & 1.50 &  \\  \hline
$ r \pm \Delta r$    & $ (-1.62 \pm 0.83)\times 10^{-13} $ &  & $ (1.25\pm 0.16)$ \\  \hline
\end{tabular}
\end{center}
\end{table}
 
\subsubsection{ Calculation of the $ \chi^{2} $ test with QED generator \cite{Berends}  }

\begin{figure}[htbp]
\vspace{0.0mm}
\begin{center}
 \includegraphics[width=8.5cm,height=7.0cm]{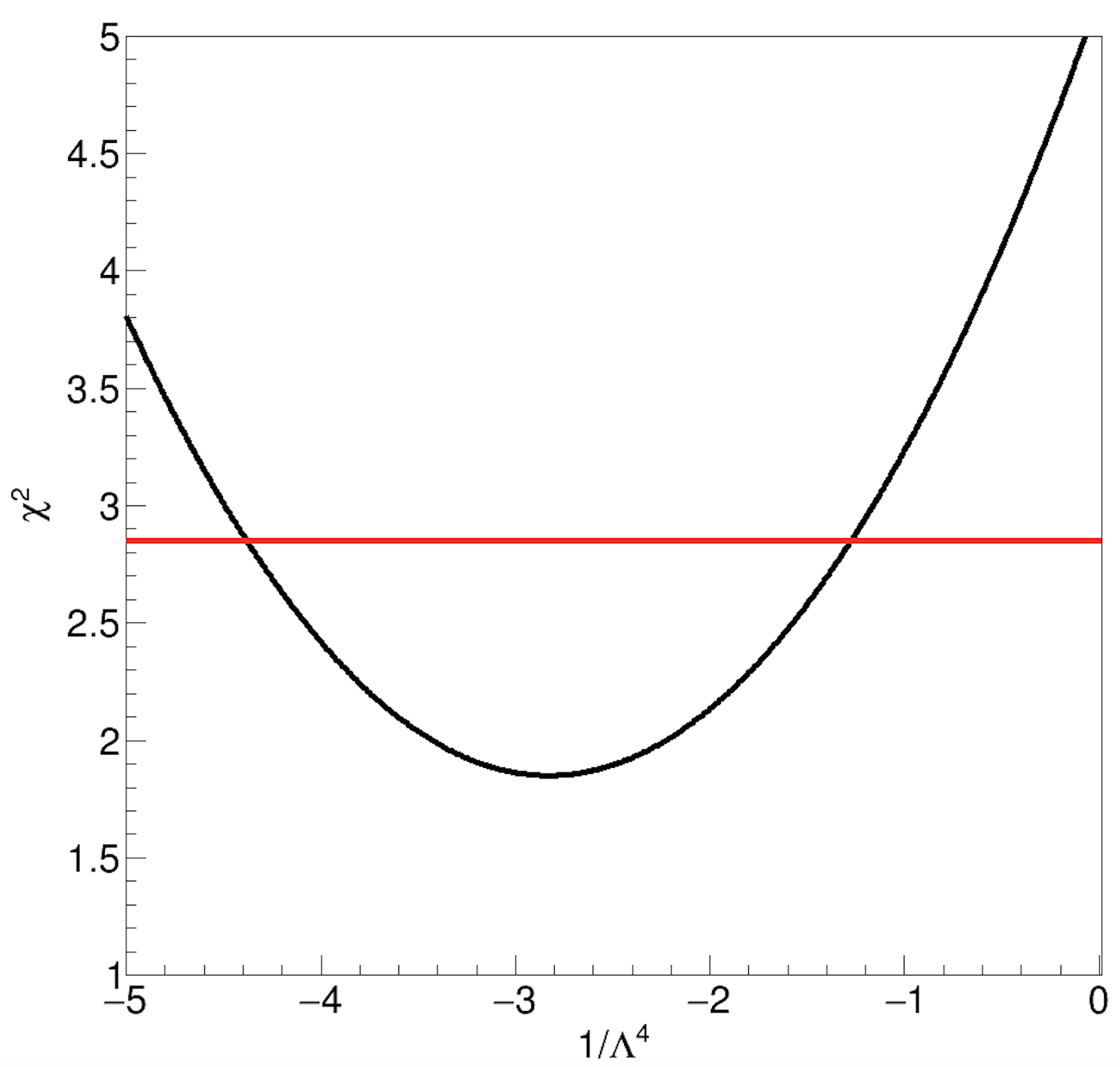}
\end{center}
\vspace{-3.0mm}
\caption{$ \chi^{2} $ test on the ratio ( \ref{CHI-1} ) of the $ \EEGG $ reaction from centre-of-mass energy 
             $ \sqrt{s} $ = 91 GeV to 207 GeV using QED data from Monte Carlo generator \cite{Berends}.
             The red line shows  one $ \sigma $ limit.  }
\label{CHIZhao}
\end{figure}

The VENUS, TOPAS, OPAL, DELPHI, ALEPH and L3 collaboration used for the QED cross section
of the $ \EEGG $ reaction the generators \cite{Berends}. As discussed in section 2.4.1 is the deviation
between BabaYaga and \cite{Berends} approximately $ 0.9 \% $ under the condition that both generators used
the same L3 parameters  ( \ref{BabaYagaL3} ). Similar to the QED BabaYaga version, the events per angular 
range are used to fit a differential cross section as function of  $ d\sigma /d\Omega \left [ pb/srad \right ]$ 
and $ \left | cos \Theta  \right | $ . The numerical parameters of this fit $ p_{1} $  to $ p_{6} $ are defined in ( \ref{QEDfit} ) 
for the 7 $ \sqrt{s} $ energies from 91.2 GeV to 200 GeV and summarised in the lower section of Table~\ref{diffpar}. 
The $ \chi^{2} $ test on the ratio ( \ref{CHI-1} ) as function of  $ \chi^{2} $ and $ 1/\Lambda ^{4} $ 
of the $ \EEGG $ reaction from centre-of-mass energy $ \sqrt{s} $ = 91.2 GeV to 200 GeV using QED data from 
Monte Carlo generator \cite{Berends} is displayed in Figure (\ref{CHIZhao}).

The minimum of the $ \chi^{2} $ test including the error of Figure (\ref{CHIZhao}) is 
$ 1/\Lambda ^{4}=-2.83_{-1.55}^{+1.56}\times 10^{-13}GeV^{-4} $. The $ \chi^{2} $ value at the minimum 
is $ \chi^{2} $ = 1.85. The fit uses 7 degrees of freedom.

After we discussed in detail, the significance $ \sigma $ and the error of the electron $ \Delta r_{e} $ in the in section 4.1.1
we concentrate in the following sections 4.1.2, 4.1.3 and 4.1.4 on the final results.

Similar to the QED - test with BabaYaga, $ 1/ \Lambda ^{4} $ in the QED - test \cite{Berends} is negative. 
A first direct approximation of the significance is $ \sigma =A/\Delta A $  = 2.83 / 1.55 = 1.83. 
In a second approximation to calculate the significance $ \sigma $ a statistical probability function $ p $ formula ( \ref{pvalue1} ) is used. 
The $ p $ value is in the discussed $ \chi^{2} $ for 7 degrees of freedom and the minimum  $ \chi^{2} $ = 1.85 equals  $ p $ = 0.032.  
Using Figure (\ref{pvaluesmall}), this is a significance of approximate $ \sigma $ = 1.9.

According to ( \ref{DIR11} ), the size $ r $ of the interaction term is $ (1.4)\times 10^{-17} $ [ cm ].
The error $ \Delta r $ is after equation (\ref{DIR18}) $ \Delta r = 0.20 \times 10^{-17} $ [ cm ]. 
The summary of all these results are given in Table~\ref{Zhao1}.

\begin{table}
 \begin{center}
 \caption{ Summary of $ \chi^{2} $ test with  generator \cite{Berends} } 
 \label{Zhao1}
 \begin{tabular}{ | l | c | r | r | }
 \hline
                    & $ (1/\Lambda )^{4} [GeV^{-4}] $     & $ \sigma $ & $ ( r \pm \Delta r ) \times 10^{-17} [cm] $  \\  \hline
direct           & $ (-2.83 \pm 1.55)\times 10^{-13} $ & 1.83 &  \\  \hline
p - value      & $ (-2.83 \pm 1.55)\times 10^{-13} $ & 1.90 &  \\  \hline
 $ r \pm \Delta r$   & $ (-2.83 \pm 1.55)\times 10^{-13} $ &  & $ (1.44\pm 0.20) $ \\  \hline
\end{tabular}
\end{center}
\end{table}

\subsubsection{ Numerical calculation of the $ \chi^{2} $ test using only direct contact term.}

The  $ \chi^{2} $ test performed with the QED BabaYaga and \cite{Berends} generator request
the calculation of the differential QED cross section and fit this data with ( \ref{QEDfit} ).

Equation ( \ref{Ratio-k} ) opens the possibility to test straight the direct contact term ( \ref{DIR3} ).
The deviation between the measured R(exp) and R(QED) is on the $ \% $ level. This opens the possibility 
to assume that the constant factors $ k_{1} $ and $ k_{2} $ are approximately equal $ k_{1} \approx k_{2} $ . 

According to Table~\ref{diffpar}, two R( QED ) data sets exist with 7 degrees of freedom from $ \sqrt{s} $ = 91.2 GeV to 200 GeV and
17 degrees of freedom from $ \sqrt{s} $ = 55 GeV to 207 GeV. Both data sets are tested.

\subsubsection{ Numerical calculation of the $ \chi^{2} $ test with 7 degrees of freedom in $ k_{1} \approx k_{2} $ approximation.}

\begin{figure}[htbp]
\vspace{-5.0mm}
\begin{center}
 \includegraphics[width=8.0cm,height=7.0cm]{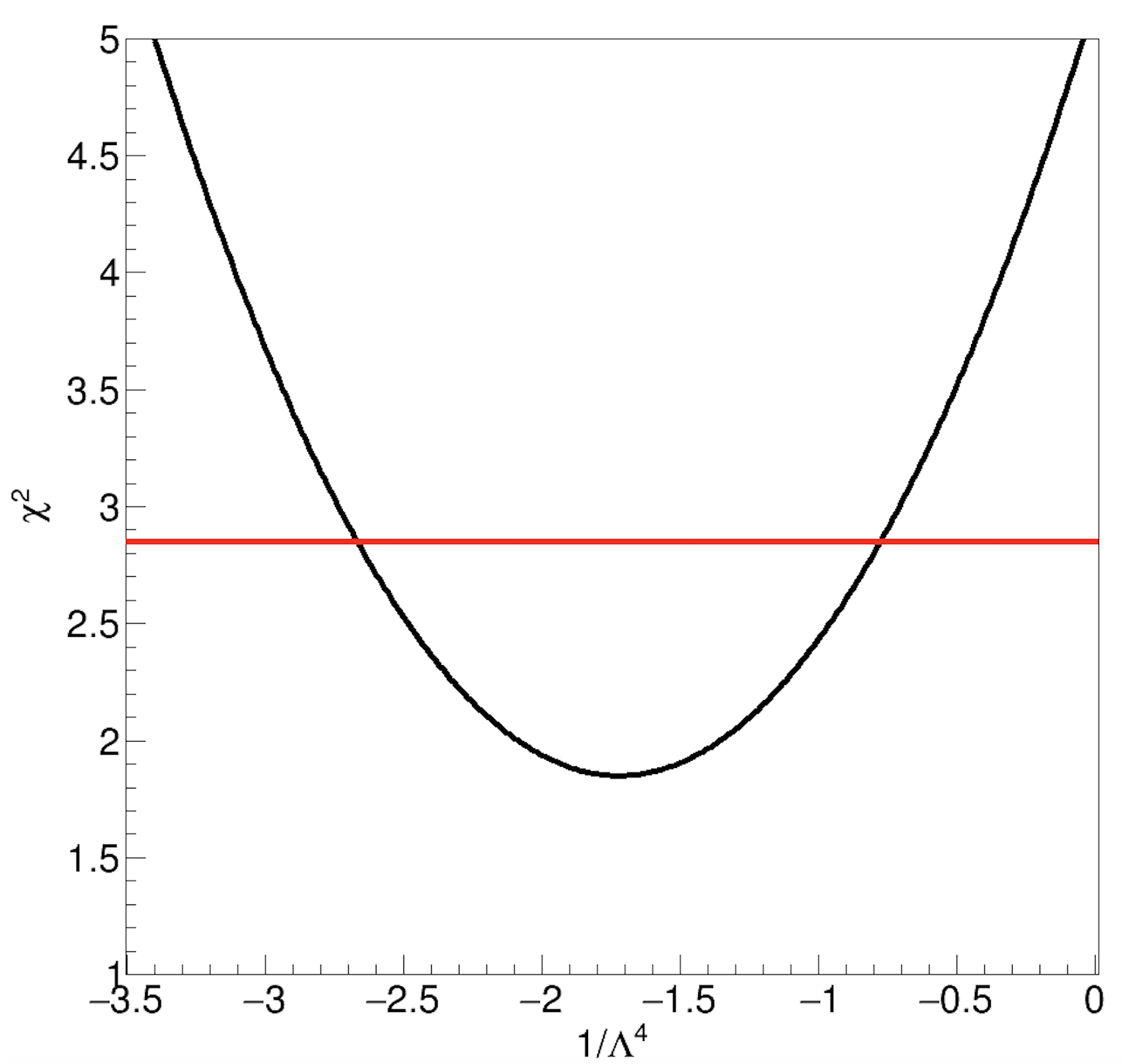}
\end{center}
\vspace{-4.0mm}
\caption{ $ \chi^{2} $ test on the ratio ( \ref{CHI-1} ) of the $ \EEGG $ reaction from centre-of-mass energy $ \sqrt{s} $ = 91.2 GeV to 200 GeV using
               7 degrees of freedom in the $ k_{1} \approx k_{2} $ approximation. The red line shows  one $ \sigma $ limit.}
\label{CHIk7d}
\end{figure}

The $ \chi^{2} $ test on the ratio ( \ref{CHI-1} ) as function of  $ \chi^{2} $ and $ 1/\Lambda ^{4} $ 
of the $ \EEGG $ reaction from centre-of-mass energy $ \sqrt{s} $ = 91.2 GeV to 200 GeV is 
displayed in Figure (\ref{CHIk7d}). 

Inserted are the 7 $ \sqrt{s} $ energies from Table~\ref{diffpar} ( Lower part ) and the according
R(exp) parameters of Table~\ref{table6}, including also  equation ( \ref{CHI-1} ),
( \ref{Ratio-11} ) and ( \ref{Ratio-k} ) under the condition $ k_{1} \approx k_{2} $.

The minimum of the $ \chi^{2} $ test including the error of Figure (\ref{CHIk7d}) is 
$ 1/\Lambda ^{4}=-1.72_{-0.95}^{+0.94}\times 10^{-13}GeV^{-4} $. The $ \chi^{2} $ value at the minimum 
is $ \chi^{2} $ = 1.85. The fit uses 7 degrees of freedom.

Similar to the QED - test BabaYaga and \cite{Berends} is $ 1/ \Lambda ^{4} $ also in this approximation  negative. 
A first direct approximation of the significance is $ \sigma =A/\Delta A $  = 2.83 / 1.55 = 1.81. 
In a second approximation to calculate the significance $ \sigma $ a statistical probability function $ p $ ( \ref{pvalue1} ) is used. 
The $ p $ value is in the discussed $ \chi^{2} $ for 7 degrees of freedom and the minimum is $ \chi^{2} $ = 1.85 equal  $ p $ = 0.032.  
Using Figure (\ref{pvaluesmall}) is this a significance of approximate $ \sigma $ = 1.8.

According to ( \ref{DIR11} ), the size $ r $ of the interaction term is $ (1.27)\times 10^{-17} $ [ cm ].
The error $ \Delta r $ is after equation (\ref{DIR18}) $ \Delta r = 0.18 \times 10^{-17} $ [ cm ].
The summary of all these results is given in Table~\ref{Zhao}

\begin{table}
 \begin{center}
 \caption{ Summary of $ \chi^{2} $ test with 7 degrees of freedom in $ k_{1} \approx k_{2} $ approximation. } 
 \label{Zhao}
 \begin{tabular}{ | l | c | r | r | }
 \hline
                    & $ ((1/\Lambda )^{4} [GeV^{-4}] $)  & $ \sigma $ & $ ( r \pm \Delta r ) \times 10^{-17} [cm] $  \\  \hline
direct           & $ (-1.72 \pm 0.95)\times 10^{-13} $ & 1.81 &  \\  \hline
p - value      & $ (-1.72 \pm 0.95)\times 10^{-13} $ & 1.80 &  \\  \hline
$ r \pm \Delta r$  & $ (-1.72 \pm 0.95)\times 10^{-13} $ &  & $ (1.27\pm 0.18) $ \\  \hline
\end{tabular}
\end{center}
\end{table}

\subsubsection{ Numerical calculation of the $ \chi^{2} $ test with 17 degrees of freedom in $ k_{1} \approx k_{2} $ approximation.}

\begin{figure}[htbp]
\vspace{0.0mm}
\begin{center}
 \includegraphics[width=8.5cm,height=7.0cm]{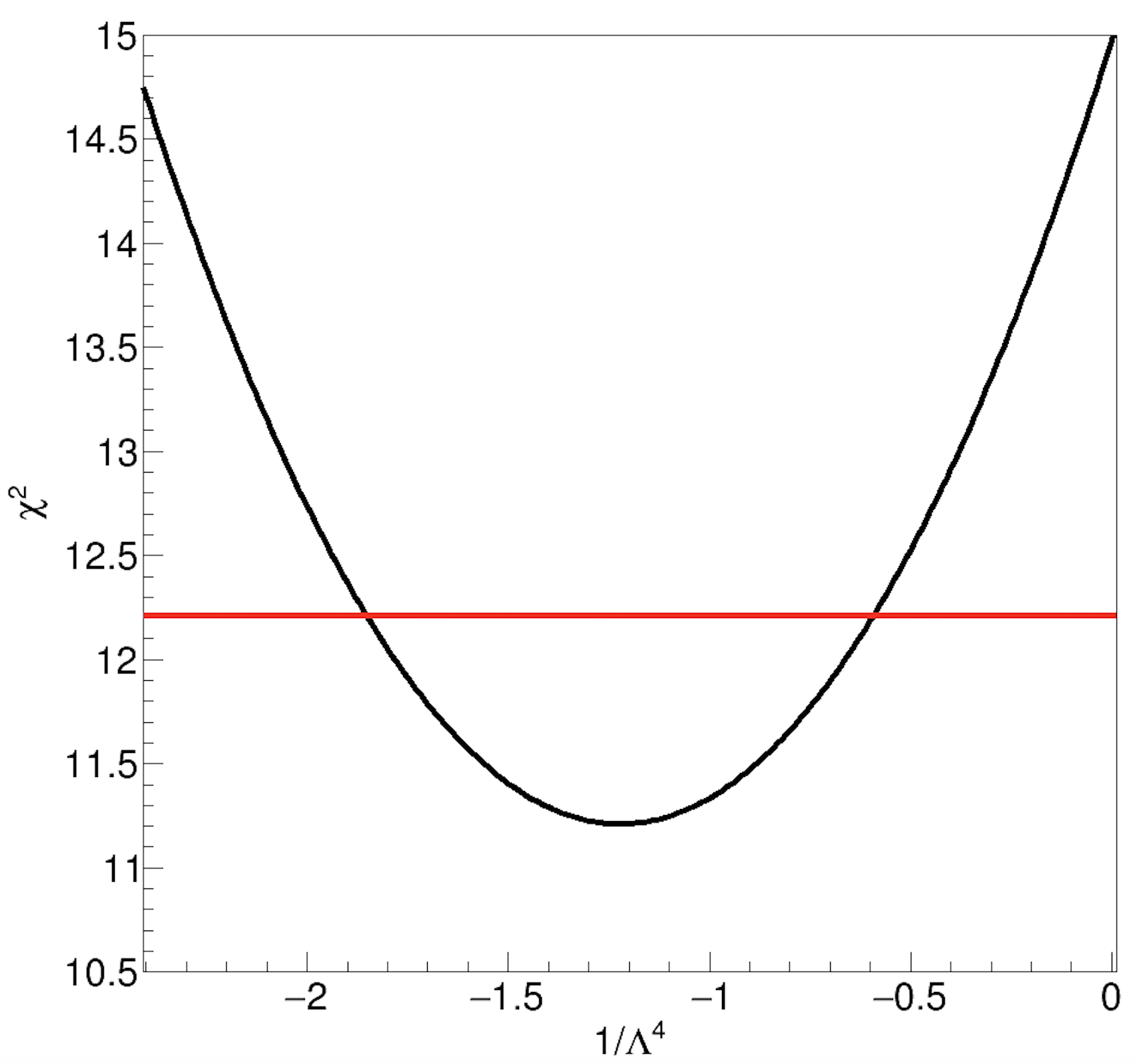}
\end{center}
\vspace{-3.0mm}
\caption{ $ \chi^{2} $ test on the ratio ( \ref{CHI-1} ) of the $ \EEGG $ reaction from centre-of-mass energy $ \sqrt{s} $ = 52 GeV to 207 GeV using
               17 degrees of freedom in the $ k_{1} \approx k_{2} $ approximation.The red line shows one  $ \sigma $ limit.}
\label{CHIk17d}
\end{figure}

The $ \chi^{2} $ test on the ratio ( \ref{CHI-1} ) as function of  $ \chi^{2} $ and $ 1/\Lambda ^{4} $ 
of the $ \EEGG $ reaction from centre-of-mass energy $ \sqrt{s} $ = 55 GeV to 207 GeV is 
displayed in Figure (\ref{CHIk17d}). Inserted in the $ \chi^{2} $ test are for the 17 degrees of freedom
and 17 $ \sqrt{s} $ energies R(exp) parameters of Table~\ref{table6} and equation ( \ref{CHI-1} ),
( \ref{Ratio-11} ) and ( \ref{Ratio-k} ) under the condition $ k_{1} \approx k_{2} $. 

The minimum of the $ \chi^{2} $ test including the error of Figure (\ref{CHIk17d}) is 
$ 1/\Lambda ^{4}=-1.22_{-0.63}^{+0.63}\times 10^{-13}GeV^{-4} $. The $ \chi^{2} $ value at the minimum 
is $ \chi^{2} $ = 11.2. The fit uses 17 degrees of freedom.

Similar to the QED - test BabaYaga and \cite{Berends} is $ 1/ \Lambda ^{4} $ also in this approximation negative. 
A first direct approximation of the significance is $ \sigma =A/\Delta A $  = 1.22 / 0.63 = 1.94. 
In a second approximation to calculate the significance $ \sigma $ a statistical probability function $ p $ ( \ref{pvalue1} ) is used. 
The $ p $ value is in the discussed $ \chi^{2} $ for 17 degrees of freedom and the minimum is $ \chi^{2} $ = 11.2 equal  $ p $ = 0.15.  
Using Figure (\ref{pvaluesmall}) is this a significance of approximate $ \sigma $ = 1.2.

According to ( \ref{DIR11} ), the size $ r $ of the interaction term is $ (1.17)\times 10^{-17} $ [ cm ]. 
The error $ \Delta r $ is after equation (\ref{DIR18}) $ \Delta r = 0.15\times 10^{-17} $ [ cm ].
The summary of all these results is given in Table~\ref{CHI17d}.

\begin{table}
 \begin{center}
 \caption{ Summary of $ \chi^{2} $ test with 17 degrees of freedom in $ k_{1} \approx k_{2} $ approximation. } 
 \label{CHI17d}
 \begin{tabular}{ | l | c | r | r | }
 \hline
                    & $ ((1/\Lambda )^{4} [GeV^{-4}] $)  & $ \sigma $ & $ ( r \pm \Delta r ) \times 10^{-17} [cm] $  \\  \hline
direct           & $ (-1.22 \pm 0.63)\times 10^{-13} $ & 1.94 &  \\  \hline
p - value      & $ (-1.22 \pm 0.63)\times 10^{-13} $ & 1.20 &  \\  \hline
$ r \pm \Delta r$  & $ (-1.22 \pm 0.63)\times 10^{-13} $ &  & $ (1.17\pm 0.15) $ \\  \hline
\end{tabular}
\end{center}
\end{table}

\subsection{ Conclusion of the four different numerical calculations of the  $ \chi^{2} $ tests. }

The four $ \chi^{2} $ tests on the ratio ( \ref{CHI-1} ) as function of  $ \chi^{2} $ and $ 1/\Lambda ^{4} $ 
of the $ \EEGG $ reaction from centre-of-mass energy $ \sqrt{s} $ = 55 GeV to 207 GeV are summarised in
Table~\ref{CHIsum}. The Table contains in Test 4.1.1 the " Calculation of the $ \chi^{2} $ test with QED BabaYaga ",
in Test 4.1.2  " Calculation of the $ \chi^{2} $ test with QED generator \cite{Berends} " , Test 4.1.4 
" Numerical calculation of the $ \chi^{2} $ test with 7 degrees of freedom in $ k_{1} \approx k_{2} $ approximation. "
and Test 4.1.5 " Numerical calculation of the $ \chi^{2} $ test with 17 degrees of freedom in $ k_{1} \approx k_{2} $ approximation. ".

All the four different $ \chi^{2} $ tests show a minimum, including a negative $ (1/\Lambda )^{4} [GeV^{-4}] $ value. 
This fact supports a negative interference effect of the direct contact term Figure (\ref{Feynman2}). 
In equation ( \ref{DIR3} ) has the term  $\delta_{new} $ negative sign. 

The sensitivities depend on the type of the $ \chi^{2} $ test and the methods of calculation. The direct calculated
sensitivity is for all $ \chi^{2} $ tests approximately stable, range from $ \sigma = 1.95 $ to $ \sigma = 1.81 $.
The sensitivity calculations via the statistics of the p - values range from $ \sigma = 1.83 $ to $ \sigma = 1.20 $.
It is important to notice, that the approximation of the k1 - k2 method is in $ \chi^{2} $ test sensitive to the energy dependence of 
the k1 - k2 factors. This dependence is not included in the $ \chi^{2} $ test and lowers the p - values for big
energy ranges, leading to  in $ \sigma = 1.20 $. Unless the $ \sigma = 1.20 $ , is the range of the sensitivities between 
direct calculated and the exact calculation of $ \sigma $ values of both methods approximately the same.
No important difference between Monte Carlo generator of BabaYage, the generator \cite{Berends} and 
the k1 - k2 approximation could be detected. 

The interaction radius $  r \pm \Delta r   $ is in the range of the statistical error $ \Delta r  $ for all $ \chi^{2} $ tests
the same. Table~\ref{CHIsum}  summarise all these information of the four $ \chi^{2} $ tests. In column 1 the method, 
column 2 the range of $ \sigma $ and in column 3 the interaction radius $  r \pm \Delta r   $ is shown. 

\begin{table}
 \begin{center}
 \caption{ Summary of the numerical calculation of the  $ \chi^{2} $ tests . } 
 \label{CHIsum}
 \begin{tabular}{ | l | c | r | r | }
 \hline
Test        & $ \sigma $ &  interaction radius $  r \pm \Delta r )  $   \\  \hline
4.1.1      & $ 1.95 - 1.50 $ & $ 1.25\pm 0.16 \times 10^{-17} [cm] ) $   \\  \hline
4.1.2      & $ 1.90 - 1.83 $ & $ 1.44\pm 0.20 \times 10^{-17} [cm] ) $   \\  \hline
4.1.4      & $ 1.81 - 1.80 $ & $ 1.27\pm 0.18 \times 10^{-17} [cm] ) $   \\  \hline
4.1.5      & $ 1.94 - 1.20 $ & $ 1.17\pm 0.15 \times 10^{-17} [cm] ) $   \\  \hline
\end{tabular}
\end{center}
\end{table}
 
It is well known statistical tests depend on the amount of data or degrees of freedom  
available for the test. This number defines finally the significance $ \sigma $. To test the differential cross 
section every angular bin of the angular distribution of the differential cross section is one degree of freedom. 
For this reason the test of the differential cross section has much higher degrees of freedom as the test of the
total cross section of the same data set.

The USTC and ETHZ collaboration published \cite{SIZE} a $ \chi^{2} $ test of the differential cross section
of the $ \EEGG $ reaction in year 2014. The measurements of the differential cross section of the $ \EEGG $ reaction from the VENUS, TOPAS, OPAL, 
DELPHI, ALEPH and L3 collaborations, collected between 1989 to 2003, are used to perform a $ \chi^{2} $ test 
to search for a finite size of an electron. Data exist between centre-of-mass energy $ \sqrt{s} $ = 55 GeV to 207 GeV 
at 17 $ \sqrt{s} $. In \cite{SIZE}, Table 4 shows the $ \chi^{2} $ test of the differential cross section
of the  $ \EEGG $ reaction. The 17  $ \sqrt{s} $ energies contain 254 degrees of freedom. 

It is interesting to notice that in \cite{SIZE}, the $ \chi^{2} $ test shows also a negative 
$ (1/\Lambda )^{4} [GeV^{-4}]  = - ( 4.05 \pm 0.73 )\times 10^{-13} $ similar to the $ \chi^{2} $ tests
of the total cross section, but with a significance of $ \sigma = 5.5 $ \cite{SIZE} ( Table~\ref{Comparison} ).

A comparison of the actual $ \chi^{2} $ test significance $ \sigma $ and the interaction radius $ r $ is shown in 
Table~\ref{Comparison}. The Table is, to guide the eye a copy of Table~\ref{CHIsum} but with the addition
in the last line including the results of the "diff-cross" is the result from \cite{SIZE}.

\begin{table}
 \begin{center}
 \caption{ Comparison of $ \chi^{2} $ tests total cross section to differential cross section. } 
 \label{Comparison}
 \begin{tabular}{ | l | c | r | r | }
 \hline
Test        & $ \sigma $ &  interaction radius $  r \pm \Delta r )  $   \\  \hline
4.1.1      & $ 1.95 - 1.50 $ & $ 1.25\pm 0.16 \times 10^{-17} [cm] ) $   \\  \hline
4.1.2      & $ 1.90 - 1.83 $ & $ 1.44\pm 0.20 \times 10^{-17} [cm] ) $   \\  \hline
4.1.4      & $ 1.81 - 1.80 $ & $ 1.27\pm 0.18 \times 10^{-17} [cm] ) $   \\  \hline
4.1.5      & $ 1.94 - 1.20 $ & $ 1.17\pm 0.15 \times 10^{-17} [cm] ) $   \\  \hline
diff-cross      & $ 5.5 $ & $  1.57\pm 0.07 \times 10^{-17} [cm] ) $   \\  \hline
\end{tabular}
\end{center}
\end{table}
 
 The difference between the $ \chi^{2} $ test of the total cross section and differential cross section
 is in the significance $ \sigma $. It confirms that the test of the differential cross section including 257 
 degrees of freedom results in a higher $ \sigma $ as 17 degrees of freedom of the total cross section measuremt. 
 Important is that the interaction radius $ ( r \pm \Delta r )  $ is for all tests in the range of the static the same.
 
 \section{Systematic errors of $ \chi^{2} $ test on the total cross section. }
 
 The default error in the ratio ( \ref{CHI-1} ) $ \Delta R(E_{i};exp;sys) $  is usually the quadratic sum of the statistical 
 error and systematic error, if both errors are independent. The data from the different groups show 
 in the total cross section the statistical error. The systematic error was not published for every group in detail.
 Following the common manner of the different collaboration in ( \ref{CHI-1} ) only the statistical error $ \Delta R(E_{i};exp) $
 is used.
 
 Systematic errors arise from the luminosity evaluation, the selection efficiency, background evaluation, the choice
 of the QED-$ \alpha^{3} $ theoretical cross section as the reference cross section, the choice of the fit procedure,
 the type of the fit parameter, and of the scattering angle in $ | cos \theta | $ for comparison between data and 
 theoretical calculations. 
  
 In Table~\ref{table6} and Figure~\ref{SIGMAtot} appears above $ \sqrt{s} $  > 91.2 GeV a small deviation in R(exp) 
 from the $ \sigma (QED)_{tot} $ cross section. The systematic error of the measured total cross section of detector$_{i} $ 
 $ \sigma(exp)_{tot}^{(det_i)} $ and total QED cross section of  $ \sigma(QED)_{tot}^{(det_i)} $  above  $  \sqrt{s} $ > 91.2 GeV, is
 for L3 \cite{L3B} ( Table 3 ) 0.10 < $ \Delta\sigma(meas.)_{sys} $ < 0.13 [ pb ] and $ \Delta\sigma(QED)_{sys} $ = 0.1 [ pb ],
 for DELPHI \cite{DELPHI} ( Table 4 , year 2000 ) 0.09 < $ \Delta\sigma(meas.)_{sys} $ < 0.14 [ pb ] and
 for OPAL \cite{OPAL2} ( Table 7 )  0.05 < $ \Delta\sigma(meas.)_{sys} $ < 0.08 [ pb ]. According to these tables 
 the $ \Delta\sigma_{sys} $ values behave statistically and the systematic error $ \Delta\sigma(meas.)_{sys} $ is
 much smaller as the statistical error $ \Delta\sigma(meas.)_{sys} < \Delta\sigma(meas.)_{stat} $.
 No change as the function of the energy of the systematic errors above $ \sqrt{s} $  > 91.2 GeV could be observed. 
 The statistically behaviour and the size of the $ \Delta\sigma(meas.)_{sys} $ excludes the possibility, that the deviation 
 of R(exp) from R(QED) could be originated from an energy $ \sqrt{s} $ behaviour of the systematic errors or the
 size of the error $ \Delta\sigma(meas.)_{sys} $.
  
\section{Discussion} 

\subsection{ Ratio plot of the total cross section  as function of the finite size of the electron.}

\begin{figure}[htbp]
\vspace{0.0mm}
\begin{center}
\includegraphics[width=8.5cm,height=6.0cm]{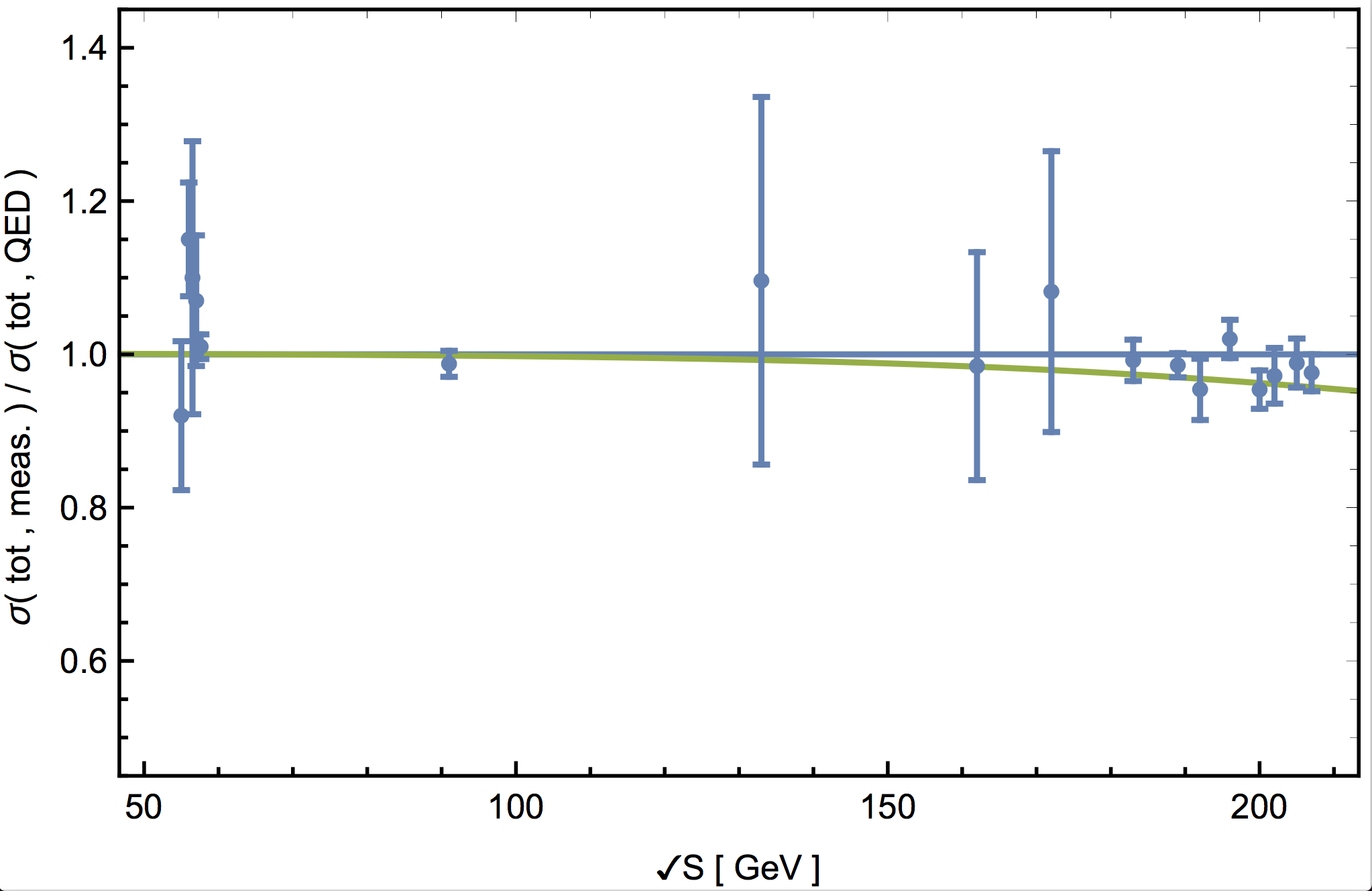}
\end{center}
\caption{ Ratio $ \sigma $ (tot,meas.) /  $ \sigma$(tot,QED)  of the $ \EEGG $ reaction of all detectors as function of  centre-of-mass energy $ \sqrt{s} $ .
                The experimental data (points), QED prediction (solid blue line) and green solid line $ \chi^{2} $ fit result of finite size of electron.}
\label{Ratio-3-final}
\end{figure}

The global $ \chi^{2} $ test under discussion uses the total cross section measured from different detectors.
To investigate the deviation from the total measured $ \EEGG $ cross section from QED total cross section, 
it is necessary to introduce an approach for a  common total cross section in the energy range from 
55 GeV < $ \sqrt{s} $ < 207 GeV shown in Figure~\ref{SIGMAtot}.
In a first view, is the agreement between the total measured cross section $ \sigma $ (tot) and the QED cross section 
$ \sigma $ (tot,QED) excellent. To test in more detail, the agreement between both cross sections, it is necessary
to study the ratio 
\newline 
 $ \sigma $ (tot, meas. ) / $ \sigma $ (tot, QED ) in Figure~\ref{Ratio-1}. 
To display the deviation from the ratio 
$ \sigma $ (tot, meas. ) / $ \sigma $ (tot, QED ) in Figure~\ref{Ratio-1}
it is possible to calculate the ratio 
\newline 
$ R(\Lambda_{6})=\sigma(QED)_{tot}^{L3}/\sigma(QED+\Lambda_{top})_{tot}^{L3} $.
This ratio is  calculated from the Monte Carlo generator, using the pure QED $ \Lambda_{6} $ from the total cross section,
in the minimum of the $ \chi^{2} $ test. The green line in Figure~\ref{Ratio-3-final} displays the effect of a finite size of
the electron generated from the  $ \chi^{2} $ test.

A deviation from $ \sigma $ (tot, meas. ) / $ \sigma $ (tot, QED) < 1.0 appears above $ \sqrt {s} $ = 180.0 GeV,
indicating, as in the $ \chi^{2} $ test, a negative interference. These findings agree with bigger statistical error bars
of the measurement from L3  \cite{L3B} ( Fig. 2 ) and DELPHI \cite{DELPHI} (2000 , Fig. 2 ). 

\subsection{ Finite size of an electron in rest. }

If the electron has a finite extension, the search with the $ \EEGG $ - reaction at high energies is, a competition between 
the Lorentz Contraction of the object and the size of the object in rest. Including the Lorentz Contraction at an $ \sqrt{s} $ = 207 GeV
the electron interaction size in rest would be with approximately 2.92 $ \times 10^{-14} $ [ m ]. For comparison, this size would be 
bigger than the charge radius of the proton 0.87 $ \times 10^{-15} $ [ m ] in rest. Under these circumstances, it seems possible to 
speculate that a charge distribution inside this electron volume exists. The effective Lagrangian of ( \ref{DIR22} ) is electromagnetic. 
The annihilation of the $ \EEGG $ - reaction would test the long-range direct contact term to the charge distribution. 

\section{Conclusion}

The total cross section of the $ \EEGG $ reaction measured from the VENUS, TOPAS, OPAL, DELPHI, 
ALEPH and L3 collaborations, was used to test the QED. 
The aim of this investigation is to prove in a global  $ \chi ^2 $ fit, that the use of the total cross section of all these data implies 
a finite size of the electron. 

First, it was necessary to discuss the theoretical framework of the calculation of the total QED cross section,
in particular the fact that an analytic precise QED cross section must be calculated via a Monte Carlo program. 

Second, all the discussed collaborations measured the total $ \EEG $ cross section but for different parameters
( Like angular range a.s.o. ). For this reason it is not possible to show a total cross section for the whole energy 
range $ \sqrt{s} $ = 55 GeV to 207 GeV. It was necessary to introduced a model for a total cross section using the 
data from the 6 collaborations. The model allows to perform a total $ \chi ^2 $ fit of all data. 

In total four different  $ \chi^{2} $ tests were used to test the finite size of the electron.
All the four different $ \chi^{2} $ tests show a minimum, including a negative $ (1/\Lambda )^{4} [GeV^{-4}] $ value. 
This fact supports a negative interference effect of the direct contact term. The maximum of the direct calculated sensitivity is
approximately $ \sigma = 1.9 $. The range of the sensitivity between all four different $ \chi^{2} $ tests is approximately the same. 
The interaction radius $  r \pm \Delta r   $ is in the range of the statistical error 
$ \Delta r  $ the same. After a common interpretation is this a " HINT " of an effect of new physics.

The measurements of the differential cross section of the $ \EEGG $ reaction from the six collaborations, 
collected between 1989 to 2003, are used to perform a global $ \chi^{2} $ test to search for a finite size of 
an electron \cite{QED-1}, \cite{SIZE}. The test detected a negative $ (1/\Lambda )^{4} [GeV^{-4}]  = - ( 4.05 \pm 0.73 )\times 10^{-13} $ 
and a radius $ r $ of the electron  of $ 1.57 \pm 0.28 \times 10^{-17} [cm] ) $. In the range of the statistical error
are this values equal to the total cross section $ \chi^{2} $ test. The differential cross section analysis confirms the total 
cross section $ \chi^{2} $ statistic. An important difference between the $ \chi^{2} $ of the differential cross section and the total cross section
is the sensitivity in the differential cross section $ \sigma $ = 5.5 and in the total cross section only $ \sigma $ = 1.9.
This difference is originated from total available data ( degrees of freedom ) from the differential cross section and the
total cross section. 

Extensive measurements and analyses are performed to search for quark and lepton compositeness in contact
interaction \cite{BHABA} in the Bhabha channel Figure~\ref{Feynman1}.  A hint for axial-vector contact interaction, in the 
data on $ e^{+} e^{-}  \rightarrow  e^{+} e^{-} ( \gamma ) $ scattering from ALEPH, DELPHI, L3 and OPAL at centre-of-mass energies
192 - 208 GeV was detected at $ \Lambda = 10.3^{+2.8}_{-1.6} $ TeV \cite{BOURILKOV}.

Depending on the experimental test, the electron exhibits two extensions. 
In the case of the $ \EEGG $ reaction, only the QED long range interaction is tested. 
The weak interaction the $ Z^{0} / \gamma $ channel Figure~\ref{Feynman1}, is suppressed by angular momentum conservation. 
In the case of the Bhabha reaction $ e^{+} e^{-}  \rightarrow  e^{+} e^{-} ( \gamma ) $, the short range weak and QED
interaction is involved. The Bhabha channel is dominating because the differential cross section in Bhabha channel is 
much bigger than in the pure QED channel. The contribution of the $ Z^{0} $ in the reaction, leads to a 
$ \Lambda = 10.3^{+2.8}_{-1.6} $ TeV, about 8 times bigger as in the $ \EEGG $ reaction.
The conclusion of these findings would be, that in the electron an outer and inner core exists. 
Some possible consequences of this result are discussed in \cite{SIZE}.

\begin{acknowledgements}

We are very grateful Andr\'{e}Rubbia, Claude Becker, Xiaolian Wang, Ziping Zhang and Zizong Xu for the many years encouraging support of the project.
In particular, we thank in memory of Hans Hofer and Hongfang Chen for their long time engagement in this experiment.

\end{acknowledgements}

\clearpage

\appendix

\part*{Appendices}

\subsection{ Virtual and soft radiative corrections of the $ \EEGG $ cross section. }

If the energy of the  photons from initial state radiation ( soft Bremsstrahlung ) 
is too small for detection $ k_{3}/|p_{+}|=k_{0}<<1 $, the reaction can be treated 
as 2-photon final state in (\ref{BORNCORR}).

\begin{equation}
\label{BORNCORR}
e^+(p_{+})+e^-(p_{-}) \rightarrow \gamma (k_{1})+\gamma (k_{2}))
\end{equation}

The equation for $ \delta _{virtual}+\delta _{soft} $ are in ( \ref{BORNCORR1} ) to ( \ref{BORNCORR4} ).

\begin{equation}
\label{BORNCORR1}
\begin{matrix}
  \delta _{soft}+\delta _{virtual}= -\frac{\alpha }{\pi } \{2(1-2v)(lnk_{0}+v)+\frac{3}{2}  \vspace{0.2cm} \\  
  -\frac{1}{3}\pi ^2 +\frac{1}{2(1+cos^2\theta )} \vspace{0.2cm} \\
\times [-4v^2(3-cos^2\theta) -8vcos^2\theta  \\ 
+4uv(5+2cos\theta +cos^2\theta ) \\ 
+4wv(5-2cos\theta ) + cos^2\theta    \\ 
-u(5-6cos\theta +cos^2\theta ) \\
-w(5+6cos\theta +cos^2)  \\ 
-2u^2(5+2cos\theta +cos^2\theta ) \\
-2w^2(5-2cos\theta +cos^2\theta ) ] \}
\end{matrix}
\end{equation}

\begin{align}
\label{BORNCORR2}
v&=\frac{1}{2}ln\left ( \frac{s}{m^2_{e}} \right )\\ 
u&=\frac{1}{2}ln\left ( \frac{2(e+cos\theta )}{m^2} \right )\\
\label{BORNCORR3}
w&=\frac{1}{2} ln\left ( \frac{2(e-cos\theta )}{m^2} \right )  \\ 
\label{BORNCORR4}
m&=\frac{m_{e}}{\left | p_{+} \right |}
\end{align}
 
The mass of the electron $ m_{e} $ is still included in this equation. The total cross section
of the two $ \gamma $ final state is in ( \ref{Stot1} )

\begin{align}
\label{Stot1}
\sigma ^{2\gamma }&=\sigma _{0}+\frac{2\alpha ^3}{s}[2(2v-1)^2lnk_{0}+\frac{4}{3}v^3+3v^2\vspace{0.2cm} \\
&+(\frac{2}{3}\pi ^2-6)v-\frac{1}{12}\pi ^2] \nonumber 
\end{align}

\subsection{ Hard radiative corrections of the $ \EEGG $ cross section. }

The soft-Bremsstrahlung photon energy is limited by a value $ k_{3}/\vert p_{1}\vert = k_{0} \ll 1 $.
If the energy of the photons from initial state radiations are above $ k_{0} $ , the process is treated 
as 3 - photon final states ( \ref{14} )

\begin{equation}
\label{14}
e^+(p_{+})+e^-(p_{-}) \rightarrow \gamma (k_{1})+\gamma (k_{2})+\gamma (k_{3})
\end{equation}

For the differential cross section of  $ \EEGG $ it is necessary to introduce two additional parameters in the phase space.
The calculation in ( \ref{HARTCORR1} ) is performed in the extreme relativistic limit \cite{Mandl}.

\begin{equation}
\label{HARTCORR1}
\frac{d\sigma }{d\Gamma _{ijk}}=\frac{d\sigma }{d\Omega_{i} d\Omega _{k} dx_{k}}=\frac{\alpha ^3}{8\pi ^2s}w_{ijk}F(i,j,k)
\end{equation}

\begin{align}
\label{HARTCORR2}
w_{ijk}&=\frac{x_{i}x_{k}}{y(z_{j})},x_{i}=\frac{k_{i0}}{\left | \vec{p_{+}} \right |}  \\
\label{HARTCORR3}
y(z_{j}) &=2e -x_{k}+x_{k}z_{j}\\ 
\label{HARTCORR4}
z_{j} &=cos(\alpha _{ik})\\
\label{HARTCORR5}
x_{l}&=\frac{E_{l}}{\left | \vec{p_{+}} \right |} 
\end{align}

\begin{align}
\label{HARTCORR6}
F(i,j,k)&=\sum_{p} \left [ -2m^2 \frac{{k_{j}}'}{k^2_{k}{k_{i}}'}-2m^2\frac{k_{j}}{{k}'^2_{k}k_{i}}+\frac{2}{k_{k}{k}'_{k}} \left (  \frac{k^2_{j}+{k}'^2_{j}}{k_{i}{k}'_{i}}\right )\right ] \nonumber \\
&=\sum_{p}M(i,j,k)
\end{align}
 
 \noindent 
 $ \alpha_{ik} $ is the angle between $ k_{i} $ and $ k_{j} $. $ P $ binds all permutations of ( i, j, k $ \in $ ( 1, 2, 3 ).
 The quantities $ k_{i} $ and $ k'_{i} $ are
  
 \begin{align}
 \label{HARTCORR7}
 k_{i}&=x_{i}(e-cos(\theta _{i})) \\
{k}'_{i}&=x_{i}(e+cos(\theta _{i}))
\end{align}

\noindent
where $ \theta _{i} $ is the angle between the momentum of the i-th photon and $ | \vec{p}_{+}| $. The
total $ 3 \times \gamma $ cross section is ( \ref{HARTCORR8} ).
 
\begin{align}
\label{HARTCORR8}
\sigma ^{3\gamma }=\frac{1}{3!}\int d\Gamma _{ijk},i,j,k\in {\{1,2,3\}}
\end{align}
 
 \noindent
 The integral runs over all phase space : $ k_{0} < x_{i} < 1 $ .
 
 For practical calculation ( \ref{HARTCORR8} ) can be approximated by an analytical approach. The photons
 get sorted after energy, $ E_{\gamma 1 } \geq E_{\gamma 2 } \geq E_{\gamma 3 } $ where 
 $ \gamma_{1} $ and $ \gamma_{2} $ are treated as annihilation photons, and $ \gamma_{3} $ as 
 hard Bremsstrahlung photon. The total cross section after integrations is ( \ref{Stot3} ) \cite{Berends} .
 
 \begin{align}
 \label{Stot3}
 \sigma ^{3 \gamma } =\frac{2\alpha ^3}{s}[3-(ln\frac{4}{m^2}-1)^2(2ln k_{0}+1)]
 \end{align}

\subsection{ Four tables needed for the numerical calculations of the $ \chi^{2} $ test. }

To calculate the differential cross section ( \ref{QEDfit} ) from the BabaYaga@nlo generator \cite {Berends} and  \cite {BABAYAGA},
the QED fit parameters $ p_{1} $  to $ p_{6} $ of the differential cross section from VENUS, TOPAS, ALEPH, DELPHI, OPAL 
normalised to L3 are given in Table~\ref{diffpar}. The parameters are sorted after the 17 $ \sqrt {s} $ energies from $ \sqrt {s} $ = 55 GeV 
to 207 GeV for the BabaYaga generator Table~\ref{diffpar} ( Upper part ) and the 7 $ \sqrt {s} $ energies from $ \sqrt {s} $ = 91.2  GeV to 
200 GeV for the generator \cite {Berends} Table~\ref{diffpar} ( Lower part ).

\begin{table*}
\caption{The QED fit parameters $ p_{1} $  to $ p_{6} $ of the differential cross section from VENUS, 
TOPAS, ALEPH, DELPHI, OPAL normalised to L3. Upper part BabaYaga, lower part \cite {Berends} }
\label{diffpar}
\begin{center}
\begin{tabular*}{\textwidth}{@{\extracolsep{\fill}}|l|r|r|r|r|r|r|@{} }
\hline
GeV& \multicolumn{1}{c|}{$ p_{1} $} & \multicolumn{1}{c|}{$ p_{2} $} & \multicolumn{1}{c|}{$ p_{3} $} & \multicolumn{1}{c|}{$ p_{4} $} 
& \multicolumn{1}{c|}{$ p_{5} $}& \multicolumn{1}{c|}{$ p_{6} $} \\
\hline
$55$     & -5.31   & 8.97  & 0.65  &  9.41  &  0.42  & -7.30  \\ \hline
$56$     & -5.68   & 9.20  & 0.66  &  9.52  &  0.31  & -7.09  \\ \hline
$56.5$   & -0.39   & 2.29  & 0.52  &  4.53  & -1.28  & -3.59  \\ \hline
$57$     & -6.81   & 9.55  & 0.65  &  9.81  & -0.10  & -6.05  \\ \hline
$57.6$   & -7.92   & 7.34  & 0.62  &  7.44  & -0.98  & -1.46  \\ \hline
$91.2 $  & -9.28   & 9.08  & 0.61  &  9.90  & -1.68  & -2.06  \\ \hline
$133$    & -9.80   & 9.42  & 0.61  &  9.99  & -1.63  & -1.89  \\ \hline
$162$    & -4.02   & 6.29  & 0.58  &  8.95  & -1.87  & -4.74  \\ \hline
$172$    & -4.17   & 6.30  & 0.58  &  8.96  & -2.13  & -4.41  \\ \hline
$183$    & -4.42   & 6.57  & 0.59  &  9.11  & -1.91  & -4.60  \\ \hline
$189$    & -3.77   & 5.84  & 0.58  &  8.82  & -2.83  & -3.84  \\ \hline
$192$    & -4.89   & 6.81  & 0.59  &  9.27  & -2.10  & -4.27  \\ \hline
$196$    & -4.63   & 6.39  & 0.58  &  9.08  & -2.37  & -3.90  \\ \hline
$200$    & -4.29   & 6.44  & 0.59  &  9.07  & -2.02  & -4.56  \\ \hline
$202$    & -4.00   & 6.22  & 0.58  &  8.93  & -2.03  & -4.60  \\ \hline
$205$    & -4.11   & 6.27  & 0.58  &  8.98  & -2.04  & -4.55  \\ \hline
$207$    & -4.05   & 6.29  & 0.58  &  8.98  & -1.95  & -4.69  \\ \hline
$ 91.2 $ & 0.95    & -0.34 & 0.08  &  1.16  & -2.70  &  1.77  \\ \hline
$ 133 $  & 0.95    & -0.31 & 0.09  &  1.19  & -2.76  &  1.81  \\ \hline
$ 161 $  & 1.04    & -0.34 & 0.13  &  0.69  & -1.95  &  1.40  \\ \hline
$ 172$   & 1.22    & -0.47 & 0.18  & -0.17  & -0.62  &  0.75  \\ \hline
$ 183 $  & 1.27    & -0.51 & 0.19  & -0.37  & -0.38  &  0.66  \\ \hline
$ 189 $  & 1.29    & -0.52 & 0.20  & -0.40  & -0.41  &  0.70  \\ \hline
$ 200 $  & 1.29    & -0.53 & 0.20  & -0.40  & -0.42  &  0.72  \\ 
\hline
\end{tabular*}
\end{center}
\end{table*}

The total experimental cross section $ \sigma(exp)_{tot}^{(det_i)} $ [pb] and the statistical error in [ pb ] from VENUS, TOPAS, ALEPH, DELPHI, OPAL and total L3 
event rate L3-N(exp) are shown in Table~\ref{EXPtot} sorted after 17 $ \sqrt {s} $ energies from $ \sqrt {s} $ = 55  GeV to 207 GeV. 

\begin{table*}
\caption{The total experimental cross section $ \sigma(exp)_{tot}^{(det_i)} $ [pb] from VENUS, TOPAS, ALEPH, DELPHI, OPAL and total L3-N(exp) event rate.}
\label{EXPtot}
\begin{center}
\begin{tabular*}{\textwidth}{@{\extracolsep{\fill}}|l|r|r|r|r|r|r|@{} }
\hline
GeV& \multicolumn{1}{c|}{$VENUS  $} & \multicolumn{1}{c|}{$TOPAS$} & \multicolumn{1}{c|}{$ALEPH$} 
& \multicolumn{1}{c|}{$DELPHI$} & \multicolumn{1}{c|}{$L3-N(exp)$}& \multicolumn{1}{c|}{$OPAL$} \\
\hline
$55$ & 46.4 $\pm$ 4.9 & & &  & &\\ \hline
$56$ &    55.8  $\pm$ 3.6 & & &  & & \\ \hline
$56.5$ &   52.5  $\pm$ 8.5 & & &  & &\\ \hline
$57$ &  50.2  $\pm$ 4.0  & & &  & & \\ \hline
$57.6$ &   & 50.2  $\pm$ 0.8  & & &  &   \\ \hline
$91.2$ &    &  & 45.13 $\pm$ 2.6 & 17.4 $\pm$ 0.8 & 1882 & 32.4 $\pm$ 2.3\\ \hline
$133$ &    & & & 9.42 $\pm$ 2.06 & &  \\ \hline
$162$ &    & & &  5.76 $\pm$ 0.87  & &  \\ \hline
$172$ &   & & & 5.55 $\pm$ 0.94  & &  \\ \hline
$183$ &    &  & & 4.27 $\pm$ 0.35  & 439  & 10.05 $\pm$ 0.43  \\ \hline
$189$ &    &  & & 4.27 $\pm$ 0.20  & 1302  & 8.79 $\pm$ 0.23 \\ \hline
$192$ &    &  & & 3.43 $\pm$ 0.43  & 193  & 9.24 $\pm$ 0.58  \\ \hline
$196$ &    &  & & 4.22 $\pm$ 0.28  & 555 & 8.43 $\pm$ 0.34  \\ \hline
$200$ &    &  & & 3.73 $\pm$ 0.25  & 424  & 7.39 $\pm$ 0.31  \\ \hline
$202$ &    &  & & 3.50 $\pm$ 0.34  & 223  & 7.88 $\pm$ 0.47 \\ \hline
$205$ &    &  & &   & 459  & 7.40 $\pm$ 0.31 \\ \hline
$207$ &    &  & & & 863 & 6.78 $\pm$ 0.23 \\ 
\hline
\end{tabular*}
\end{center}
\end{table*}

The total QED cross section $ \sigma(QED)_{tot}^{(det_i)} $ [pb] from 
\newline
VENUS, TOPAS, ALEPH, DELPHI, OPAL and L3 are shown in 
Table~\ref{QEDtot}. The Table~\ref{QEDtot} is sorted after the 17 $ \sqrt {s} $ energies from $ \sqrt {s} $ = 55  GeV to 207 GeV.
As discussed in chapter 3.1 and 3.2, the 17 lines of the Table~\ref{QEDtot} include separate information if 
only one detector or more as one detector is involved in the calculation of the total QED cross section.

If only one detector is involved in the lines  $ VENUS - L3_{norm} $ from $ \sqrt {s} $ = 55 GeV to 57 GeV.
Recorded left side is the VENUS QED total cross section in [ pb ] and right side the L3 normalised total cross section in [ pb ].
The same situation counts for $ TOPAS - L3_{norm} $ at $ \sqrt {s} $ = 57.6 GeV and $ DELPHI-(L3_{norm})$ from
$ \sqrt {s} $ = 133 GeV to 172 GeV. Recorded left side is the QED total cross section in [ pb ] and right side the L3 normalised 
total cross section in [ pb ].

If mores as one detector is involved at $ \sqrt {s} $ = 91.2 GeV and from 183 GeV to 207 GeV. In this case the QED total cross 
section from the detector in [ pb ] is recorded ( Not normalised to L3 ) with the exception of $L3-N_{QED} $. Left side of the line shows the total 
QED cross section of L3 and right side the total event rate of the L3 at this $ \sqrt {s} $ energy.

\begin{table*}
\caption{The total QED cross section $ \sigma(QED)_{tot}^{(det_i)} $ [pb] from VENUS, TOPAS, ALEPH, DELPHI, OPAL and total L3-$ N(QED)^{L3} $ event rate.}
\label{QEDtot} 
\begin{center}
\begin{tabular*}{\textwidth}{@{\extracolsep{\fill}}|l|r|r|r|r|r|r|@{} }
\hline
GeV& \multicolumn{1}{c|}{$VENUS-L3_{norm} $} & \multicolumn{1}{c|}{$TOPAS-L3_{norm} $} & \multicolumn{1}{c|}{$ALEPH$} 
& \multicolumn{1}{c|}{$DELPHI-L3_{norm}$} & \multicolumn{1}{c|}{ $ L3-N(QED)^{L3} $} & \multicolumn{1}{c|}{$OPAL$} \\
\hline
$55$ &   50.4 - 136  & & &  & &\\ \hline
$56$ &   48.5 - 131  & & &  & & \\ \hline
$56.5$ & 47.7 - 129  & & &  & &\\ \hline
$57$ &   46.9 - 127  & & &  & & \\ \hline
$57.6$ &    & 49.7 - 124 & &  &  &   \\ \hline
$91.2$ &    &  & 42.8 & 18.3  & 50.9 - 1890 & 32.0 \\ \hline
$133$ &     & & & 8.59 - 24.2 & &  \\ \hline
$162$ &     & & & 5.85 - 16.3 & &  \\ \hline
$172$ &     & & & 5.13 - 14.5 & &  \\ \hline
$183$ &    &  & & 4.57 & 12.7 - 457  & 9.32  \\ \hline
$189$ &    &  & & 4.28 & 11.9 - 1360 & 8.74 \\ \hline
$192$ &    &  & & 4.15 & 11.6 - 208  & 8.47  \\ \hline
$196$ &    &  & & 3.98 & 11.1 - 574  & 8.13  \\ \hline
$200$ &    &  & & 3.82 & 10.6 - 450  & 7.81  \\ \hline
$202$ &    &  & & 3.74 & 10.4 - 234  & 7.65  \\ \hline
$205$ &    &  & &   & 10.1 - 469     & 7.42 \\ \hline
$207$ &    &  & & & 9.90 - 845 & 7.29 \\ 
\hline
\end{tabular*}
\end{center}
\end{table*}

The luminosity used from the VENUS, TOPAS, ALEPH, DELPHI, L3 and OPAL experiment is shown in Table~\ref{LUMI}. 
In the Table~\ref{LUMI} is after the value of the luminosity always the reference given the values a taken from.
The same references are also used for Table~\ref{EXPtot} and Table~\ref{QEDtot} .

\begin{table*}
\caption{The luminosity used from the VENUS, TOPAS, ALEPH, DELPHI, L3 and OPAL experiment plus references the total cross section is published }
\label{LUMI}
\begin{center}
\begin{tabular*}{\textwidth}{@{\extracolsep{\fill}}|l|r|r|r|r|r|r|@{} }
\hline
GeV& \multicolumn{1}{c|}{$VENUS$} & \multicolumn{1}{c|}{$TOPAS$} & \multicolumn{1}{c|}{$ALEPH$} & \multicolumn{1}{c|}{$DELPHI$} & \multicolumn{1}{c|}{$L3$}& \multicolumn{1}{c|}{$OPAL$} \\
\hline
$55$ & 2.34 $pb^{-1}$ \cite{VENUS}& & &  & &\\ \hline
$56$ &    5.18 $pb^{-1}$ \cite{VENUS} & & &  & & \\ \hline
$56.5$ &   0.86 $pb^{-1}$ \cite{VENUS} & & &  & &\\ \hline
$57$ &    3.70 $pb^{-1}$ \cite{VENUS} & & &  & & \\ \hline
$57.6$ &   & 52.26 $pb^{-1}$ \cite{TOPAS} & & &  &   \\ \hline
$91$ &    &  & 8.5 $pb^{-1}$ \cite{ALEPH} & 36.9 $pb^{-1}$ \cite{DELPHI}& 64.6 $pb^{-1}$ \cite{L3A}& 7.2 $pb^{-1}$ \cite{OPAL1}\\ \hline
$133$ &    & & & 5.92 $pb^{-1}$ \cite{DELPHI} & &  \\ \hline
$162$ &    & & & 9.58 $pb^{-1}$ \cite{DELPHI} & &  \\ \hline
$172$ &   & & & 9.80 $pb^{-1}$ \cite{DELPHI} & &  \\ \hline
$183$ &    &  & & 52.9 $pb^{-1}$ \cite{DELPHI}& 54.8 $pb^{-1}$ \cite{L3B}& 55.6 $pb^{-1}$ \cite{OPAL2}\\ \hline
$189$ &    &  & & 151.9 $pb^{-1}$ \cite{DELPHI}& 175.3$pb^{-1}$ \cite{L3B}& 181.1 $pb^{-1}$ \cite{OPAL2}\\ \hline
$192$ &    &  & & 25.1$pb^{-1}$ \cite{DELPHI}& 28.8 $pb^{-1}$ \cite{L3B}& 29.0 $pb^{-1}$ \cite{OPAL2}\\ \hline
$196$ &    &  & & 76.1 $pb^{-1}$ \cite{DELPHI}& 82.4$pb^{-1}$ \cite{L3B}& 75.9 $pb^{-1}$ \cite{OPAL2}\\ \hline
$200$ &    &  & & 82.6 $pb^{-1}$ \cite{DELPHI}& 67.5 $pb^{-1}$ \cite{L3B}& 78.2 $pb^{-1}$ \cite{OPAL2}\\ \hline
$202$ &    &  & & 40.1 $pb^{-1}$ \cite{DELPHI}& 35.9 $pb^{-1}$ \cite{L3B}& 36.8 $pb^{-1}$ \cite{OPAL2}\\ \hline
$205$ &    &  & & & 74.3 $pb^{-1}$ \cite{L3B}& 79.2 $pb^{-1}$ \cite{OPAL2}\\ \hline
$207$ &    &  & & & 138.1 $pb^{-1}$ \cite{L3B}& 136.5 $pb^{-1}$ \cite{OPAL2}\\ 
\hline
\end{tabular*}
\end{center}
\end{table*}

\subsection{ Equations for the calculation of significance and errors of the  $ \chi^{2} $ test. }

Using the fit program Minuit \cite{MINUIT} nearby the minimum of the parameter $ \Lambda $ an error $ \Delta \Lambda $ 
is calculated. As shown in Figure (\ref{CHIBhabha}) to Figure (\ref{CHIk17d}) the program sets a limit ( Red line ) to allow to set  
 $ \Delta \Lambda $ values. The crossing point of the $ \chi^{2} $ function with the red line gives the size $ \Delta \Lambda $ of  on the 
 $ 1/\Lambda ^{4} $ axis. It is common to calculate the significance $ \sigma $ of a statistical test in a first approximation by dividing the value of a fit
parameter $ A $ by the error bar of $ \Delta A $ like $ \sigma =A/\Delta A $. 

In addition, it is possible to use statistic theory ( \cite{WIPI11} , \cite{SIGNI} ) to calculate the significance $ \sigma $.  
For the $ \chi^{2} $ test, a statistical function f = ( NUMBER of degrees of freedom , Minimum of the $ \chi^{2} $ )
is used to calculate a $ p $ value. This $ p $ value allows also to calculate the significance $ \sigma $. 

\subsubsection{ Equations for the calculation of the significance of a $ \chi^{2} $ test. }

To calculate the statistical significance p of a  $ \chi^{2} $ test, a statistical probability function of 
f = ( NUMBER of degrees of freedom $ ( l ) $ , Minimum of the $ \chi^{2} $ $ T_{obs} $ ) is used.
The integral of this function f integrated from $ \chi^{2} $ = 0 to the minimum in $ \chi^{2} $ = $ T_{obs} $.
After equation 17 in \cite{SIGNI} is the p - value $ p = P(T<T_{obs}|l) $ ( \ref{pvalue1} ).

\begin{align}
 \label{pvalue1}
 p=\int_{0}^{T_{obs}}d\chi ^{2}\frac{1}{2^{l/2}\Gamma (l/2)}e^{-\chi ^{2}/2}\left ( \chi ^{2} \right )^{-1+l/2}
 \end{align}

The gamma function $ \Gamma (\alpha ) $ ( \ref{pvalue2} )includes the parameter $ \alpha $ = $ l / 2 $ with $ l $ the degrees of freedom.

\begin{align}
\label{pvalue2}
\Gamma (\alpha )=\int_{0}^{\infty }y^{\alpha -1}e^{-y}dy
\end{align}

The $ p $ value is connected including the $ Erf $ function to the number of STANDART DEVIATIONS  $ \sigma $. The parameter 
$ sg $ in ( \ref{pvalue3} ) is $ sg = \sigma  $.

\begin{align}
\label{pvalue3}
p=1-Erf(sg/\sqrt{2})/2
\end{align}

\begin{figure}[htbp]
\vspace{-10.0mm}
\begin{center}
 \includegraphics[width=9.0cm,height=7.0cm]{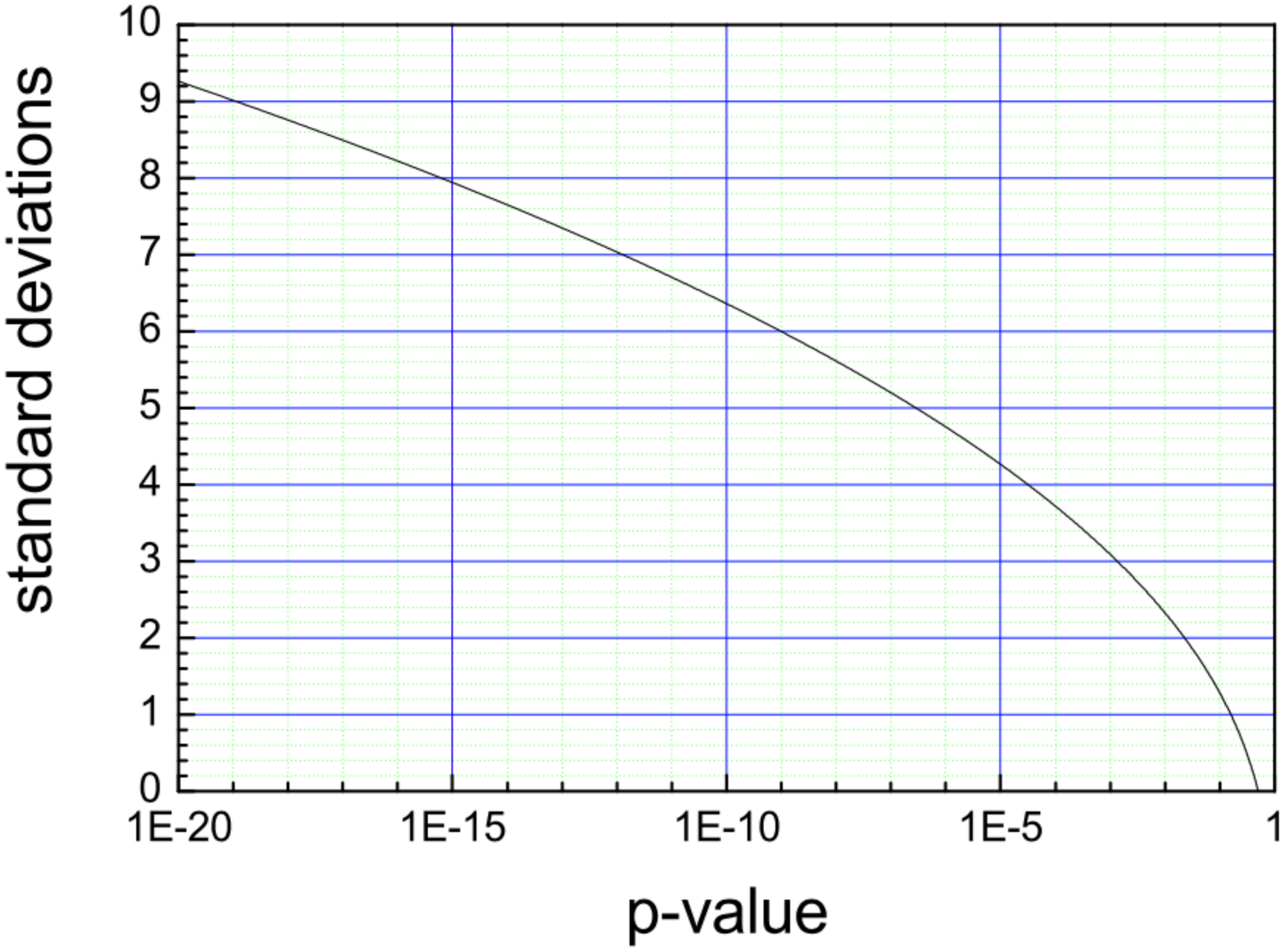}
\end{center}
\vspace{-5.0mm}
\caption{ The p - value ( \ref{pvalue1} ) as function of STANDART DEVIATIONS $ sg=\sigma $ = 1 to 10 ( \ref{pvalue3} )   }
\label{pvaluebig}
\end{figure}

In Figure (\ref{pvaluebig})  the function of STANDART DEVIATIONS between $ sg=\sigma $ = 1 to 10   via p - value are displayed  \cite{SIGNI}.

\begin{figure}[htbp]
\vspace{0.0mm}
\begin{center}
 \includegraphics[width=9.5cm,height=7.0cm]{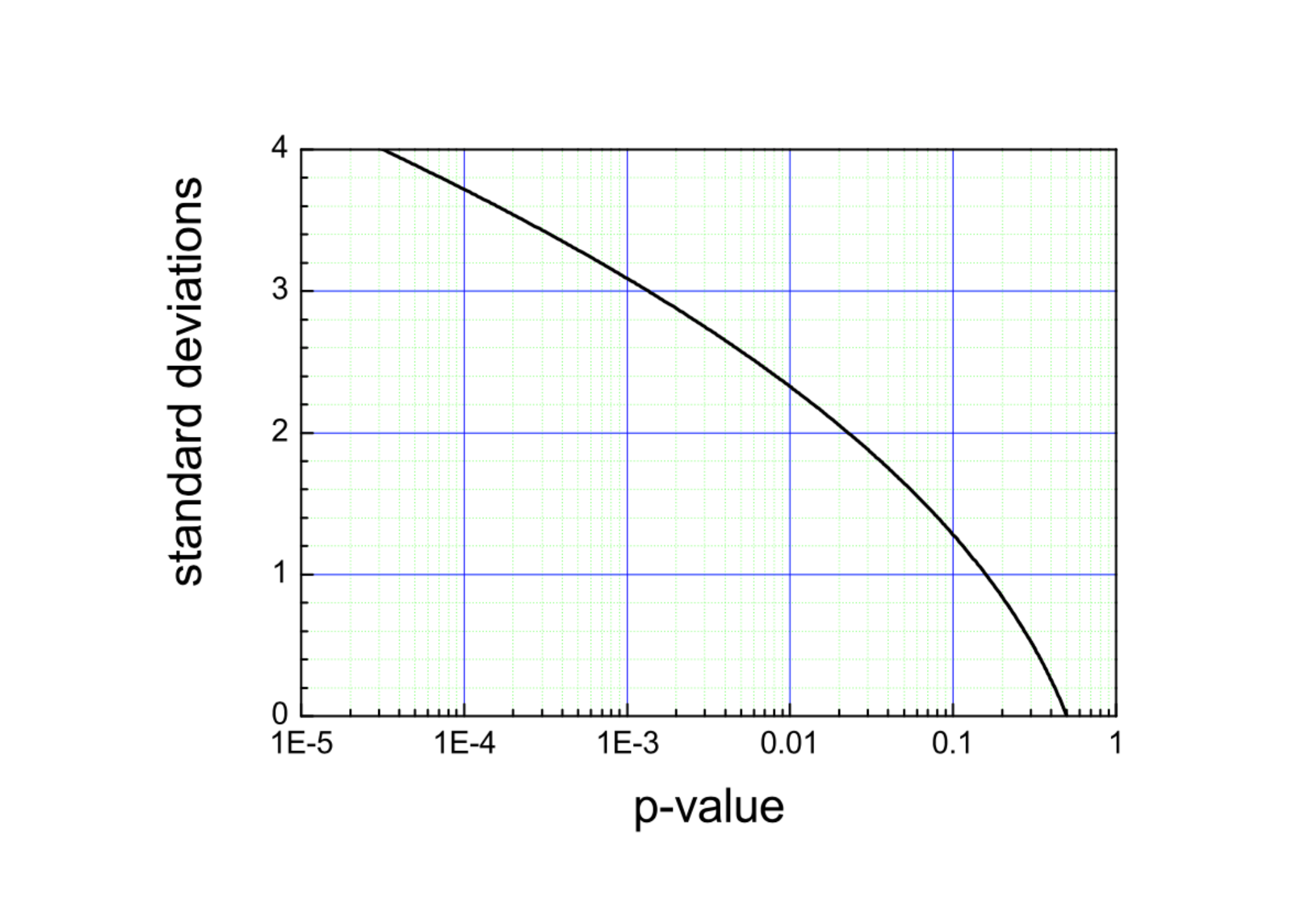}
\end{center}
\vspace{-10.0mm}
\caption{The p - value ( \ref{pvalue1} ) as function of STANDART DEVIATIONS $ sg= \sigma $ = 1 to 4 ( \ref{pvalue3} )  }
\label{pvaluesmall}
\end{figure}

In Figure (\ref{pvaluesmall})  the function of STANDART DEVIATIONS between $ sg=\sigma $ = 1 to 4   via p - value are displayed  \cite{SIGNI}.

\subsubsection{ The radius $ r_{e} $ and the error $ \Delta r_{e} $ of the interaction size of the electron in the  $ \chi^{2} $ test. }

The minimum of the fit parameter $ 1/\Lambda ^{4} $ in the $ \chi^{2} $ test is a measure
for the interaction size $ r_{e} $ of the electron (\ref{DIR11}) (\ref{DIR12}).

\begin{align}
\label{DIR12}
r_{e}=\frac{\hbar c}{\Lambda }
\end{align}

The minimum in the $ \chi^{2} $ test is given in units y = $ \frac{1}{\Lambda ^{4}}=y.yy_{-\Delta y.yy}^{+\Delta y.yy}\times GeV^{-4} $. 
For example in accordance with Figure (\ref{CHIk7d}) y = $ \frac{1}{\Lambda ^{4}}=-1.72_{-0.95}^{+0.94}\times 10^{-13}GeV^{-4} $.
To develop the parameter $ \Lambda $ as function of y in (\ref{DIR13}) 

\begin{align}
\label{DIR13}
\Lambda =\left ( \frac{1}{y} \right )^{1/4}
\end{align}

and use (\ref{DIR12}) allows to calculate the size $ r_{e} $ of the interaction radius.
In the discussed example is the interaction radius $ r_{e} = (1.271)\times 10^{-17} $ [ cm ].

The error propagation of $ \Delta r $ is calculated in (\ref{DIR14}).

\begin{align}
\label{DIR14}
\Delta r=\hbar \times c\times \frac{dr}{d\Lambda }\times \Delta \Lambda =-\frac{ \hbar \times c}{\Lambda ^{2}}\times \Delta\Lambda 
\end{align}

In (\ref{DIR14}) is $ \hbar $ the Planck constant, $ c $ the velocity of light, $ \Lambda $ the cut of parameter from the
$ \chi^{2} $ test and the error $ \Delta\Lambda $. The error calculation of the parameter $ \Lambda $ is 
not a linear function of the minimum in $ \chi^{2} $. 

Using y = $ \frac{1}{\Lambda ^{4}} $, calculating the error of y $ \Delta y $ equation (\ref{DIR15}), the error of $ \Lambda $
$ \Delta\Lambda $equation (\ref{DIR16}) and the first differential quotient $ dy/ d\Lambda $ equation (\ref{DIR17}) it
is possible calculate the error of  $ \Delta\Lambda $ in equation (\ref{DIR18}).

\begin{align}
\label{DIR15}
\Delta y=\frac{dy}{d\Lambda }\Delta \Lambda 
\end{align}

\begin{align}
\label{DIR16}
\Delta \Lambda =\frac{\Delta y}{(dy/d\Delta )}
\end{align}

\begin{align}
\label{DIR17}
dy/d\Delta =-4\Lambda ^{-5}
\end{align}

\begin{align}
\label{DIR18}
\Delta \Lambda =-\frac{1}{4}\Delta y\Lambda ^{5}
\end{align}

In the discussed example Figure (\ref{CHIk7d}), the error would be $ \Delta \Lambda =  $  214.41[ GeV ].
Inserting this value in (\ref{DIR14}) would result in a $ \Delta r  = 0.1755 \times 10^{-17} $ [cm] .

\subsubsection{ Conclusion equations for the calculation of significance and errors of the  $ \chi^{2} $ test. }
 
The equations ( \ref{pvalue1} ) to ( \ref{pvalue3} ) given an overview to calculate the significance $ \sigma $ of the
$ \chi^{2} $ - test. The significance is including the middle of the ranges from $ 1.6 < \sigma < 1.9 $ a hint of new physics.
The error $ \Delta r_{e} $ of the interaction size of the electron in the  $ \chi^{2} $ - test is
described in the ( \ref{DIR14} ) to ( \ref{DIR18} ). It is interesting to notice that according Table~\ref{Comparison} 
two independent $ \chi^{2} $ tests of the differential cross section and the total cross section result in the range of the
error to the same interaction radius $ \Delta r_{e} $.

\end{document}